# Hyperbolic Metamaterials

Igor I. Smolyaninov

*Department of Electrical and Computer Engineering, University of Maryland, College Park, MD 20742, USA*

**Hyperbolic metamaterials were originally introduced to overcome the diffraction limit of optical imaging. Soon thereafter it was realized that hyperbolic metamaterials demonstrate a number of novel phenomena resulting from the broadband singular behavior of their density of photonic states. These novel phenomena and applications include super resolution imaging, new stealth technologies, enhanced quantum-electrodynamic effects, thermal hyperconductivity, superconductivity, and interesting gravitation theory analogues. Here we briefly review typical material systems, which exhibit hyperbolic behavior and outline important applications of hyperbolic metamaterials.**

**Table of Content**

## 1. Hyperbolic metamaterial geometries and basic properties

Hyperbolic metamaterials are extremely anisotropic uniaxial materials, which behave like a metal in one direction and like a dielectric in the orthogonal direction. Originally introduced to overcome the diffraction limit of optical imaging [1-2], hyperbolic metamaterials demonstrate a number of novel phenomena resulting from the broadband singular behavior of their density of photonic states [3], which range from super resolution imaging [2,4,5] to enhanced quantum-electrodynamic effects [6,7,8], new stealth technology [9], thermal hyperconductivity [10], high Tc superconductivity [11,12], and interesting gravitation theory analogues [3, 13-17]. In the early days of metamaterial research it was believed that only artificially structured materials may exhibit hyperbolic properties. However, later on it was realized that quite a few natural materials may exhibit hyperbolic properties in some frequency ranges [11,18]. Moreover, even the physical vacuum may exhibit hyperbolic metamaterial properties if subjected to a very strong magnetic field [19].

Basic electromagnetic properties of hyperbolic metamaterials may be understood by considering a non-magnetic uniaxial anisotropic material with dielectric permittivities $\varepsilon_x=\varepsilon_y=\varepsilon_1$ and $\varepsilon_z=\varepsilon_2$. Any electromagnetic field propagating in this material may be expressed as a sum of ordinary and extraordinary contributions, each of these being a sum of an arbitrary number of plane waves polarized in the ordinary ($E_z$=0) and extraordinary ($E_z\neq 0$) directions. Let us define a "scalar" extraordinary wave function as $\varphi=E_z$ so that the ordinary portion of the electromagnetic field does not contribute to $\varphi$. Maxwell equations in the frequency domain results in the following wave equation for $\varphi_\omega$ if $\varepsilon_1$ and $\varepsilon_2$ are kept constant inside the metamaterial [3]:

$$\frac{\omega^2}{c^2}\varphi_\omega = -\frac{\partial^2\varphi_\omega}{\varepsilon_1\partial z^2} - \frac{1}{\varepsilon_2}\left(\frac{\partial^2\varphi_\omega}{\partial x^2}+\frac{\partial^2\varphi_\omega}{\partial y^2}\right) \tag{1.1}$$

While in ordinary elliptic anisotropic media both $\varepsilon_1$ and $\varepsilon_2$ are positive, in hyperbolic metamaterials $\varepsilon_1$ and $\varepsilon_2$ have opposite signs. These metamaterials are typically composed of multilayer metal-dielectric or metal wire array structures, as shown in Fig. 1. The opposite signs of $\varepsilon_1$ and $\varepsilon_2$ lead to two important consequences. For extraordinary waves in a usual uniaxial dielectric metamaterial, the dispersion law

$$\frac{k_{xy}^2}{\varepsilon_2} + \frac{k_z^2}{\varepsilon_1} = \frac{\omega^2}{c^2} \tag{1.2}$$

describes an ellipsoid in the wave momentum (k-) space (which reduces to a sphere if $\varepsilon_1 = \varepsilon_2$, as shown in Fig.2(a)). The absolute value of the k-vector in such a material is finite which leads to the usual diffraction limit on resolution of regular optics. The phase space volume enclosed between two such equi-frequency surfaces is also finite, corresponding to a finite density of photonic states. However, when one of the components of the dielectric permittivity tensor is negative, eq.(1.2) describes a hyperboloid in the phase space (Fig. 2(b)). As a result, the absolute value of the k-vector is not limited, thus enabling super-resolution imaging with hyperbolic metamaterials. Moreover, the phase space volume between two such hyperbolloids (corresponding to different values of frequency) is infinite (see Fig. 2(c)). The latter divergence leads to an infinite density of photonic states. While there are many mechanisms leading to a singularity in the density of photonic states, this one is unique as it leads to the infinite value of the density of states for every frequency where different components of the dielectric permittivity have opposite signs. It is this behavior which lies in the heart of the robust performance of hyperbolic metamaterials: while disorder can change the magnitude of the dielectric permittivity components, leading to a deformation of the corresponding hyperboloid in the phase (momentum) space, it will remain a hyperboloid and will therefore still support an infinite density of states. Such effective medium description will eventually fail at the point when the wavelength of the propagating mode becomes comparable to the size of the hyperbolic metamaterial unit cell *a*, introducing a natural wave number cut-off:

$$k_{\max} = 1/a \tag{1.3}$$

Depending on the metamaterial design and the fabrication method used, the unit cell size in optical metamaterials runs from *a*~10 nm (in semiconductor [20] and metal-dielectric layered structures [6]) to *a*~100 nm (in nanowire composites [21],[22]). Since the “hyperbolic" ehnancement factor in the density of states [3] scales as

$$\rho(\omega) = \rho_0(\omega)\left(\frac{k_{\max}}{\omega / c}\right)^3 \tag{1.4}$$

where $\rho_0 \sim \omega^2$ is the free-space result, even with the cut-off taken into account, the hyperbolic singularity leads to the optical density of states enhancement by a factor of $10^3$-$10^5$. Physically, the enhanced photonic density of states in the hyperbolic metamaterials originates from the waves with high wave numbers that are supported by the system. Such propagating modes that can achieve X-ray wavelengths at optical frequencies, do not have an equivalent in ”regular" dielectrics where $k \le \sqrt{\varepsilon}\omega / c$. Since each of these waves can be thermally excited, a hyperbolic metamaterial shows a dramatic enhancement in the radiative transfer rates.

As has been mentioned above, artificial hyperbolic metamaterials are typically composed of multilayer metal-dielectric or metal wire array structures, as shown in Fig. 1. For the multilayer geometry the diagonal components of the metamaterial permittivity can be calculated based on the Maxwell-Garnett approximation as follows:

$$\varepsilon_1 = \varepsilon_{xy} = n\varepsilon_m + (1-n)\varepsilon_d\,, \qquad \varepsilon_2 = \varepsilon_z = \frac{\varepsilon_m \varepsilon_d}{(1-n)\varepsilon_m + n\varepsilon_d} \tag{1.5}$$

where n is the volume fraction of the metallic phase, and $\varepsilon_m<0$ and $\varepsilon_d>0$ are the dielectric permittivities of the metal and dielectric, respectively [23]. The validity of Maxwell-Garnett approximation has been clearly demonstrated in ref.[23]. Analytical calculations based on the Maxwell-Garnett approximation performed for periodic array of metal nanolayers were confronted with exact numerical solutions of Maxwell equations. Excellent agreement between numerical simulations and analytical results was demonstrated. The Maxwell-Garnett approximation may also be used for a wire array metamaterial structure [23]. In this case the diagonal components of the permittivity tensor may be obtained as

$$\varepsilon_1 = \varepsilon_{x,y} = \frac{2n\varepsilon_m\varepsilon_d + (1-n)\varepsilon_d(\varepsilon_d + \varepsilon_m)}{(1-n)(\varepsilon_d + \varepsilon_m) + 2n\varepsilon_d}\;, \qquad \varepsilon_2 = \varepsilon_z = n\varepsilon_m + (1-n)\varepsilon_d \tag{1.6}$$

Since both $\varepsilon_m$ and $\varepsilon_d$ depend on frequency, the frequency regions where $\varepsilon_1$ and $\varepsilon_2$ have opposite signs may be typically found for both multilayer and wire array geometries. Depending on the actual signs of $\varepsilon_1$ and $\varepsilon_2$, the phase space shape of the hyperbolic dispersion law may be either a one-sheet ($\varepsilon_2>0$ and $\varepsilon_1<0$, see Fig.2(b)) or two-sheet ($\varepsilon_2<0$ and $\varepsilon_1>0$, see Fig.2(d)) hyperboloid. However, in both cases the k-vector is not limited, and the photonic density of states exhibits broadband divergent behavior.

We should also note that it is relatively easy to emulate various 3D hyperbolic metamaterial geometries by planar plasmonic metamaterial arrangements. While rigorous description of such metamaterials in terms of Diakonov surface plasmons may be found in ref. [24], qualitative analogy between 3D and 2D metamaterials may be explained as follows. A surface plasmon (SP) propagating over a flat metal-dielectric interface may be described by its well-known dispersion relation shown in Fig.3.

$$k_p = \frac{\omega}{c}\left(\frac{\varepsilon_d \varepsilon_m}{\varepsilon_d + \varepsilon_m}\right)^{1/2} \tag{1.7}$$

where metal layer is considered to be thick, and $\varepsilon_m(\omega)$ and $\varepsilon_d(\omega)$ are the frequency-dependent dielectric constants of the metal and dielectric, respectively [25]. Thus, similar to the 3D case, we may introduce an effective 2D dielectric constant $\varepsilon_{2D}$, which characterizes the way in which SPs perceive the dielectric material deposited onto the metal surface. By requiring that $k_p=\varepsilon_{2D}^{1/2}\omega/c$, we obtain

$$\varepsilon_{2D} = \left(\frac{\varepsilon_d \varepsilon_m}{\varepsilon_d + \varepsilon_m}\right) \tag{1.8}$$

Equation (1.8) makes it obvious that depending on the plasmon frequency, SPs perceive the dielectric material bounding the metal surface (for example a PMMA layer) in drastically different ways. At low frequencies $\varepsilon_{2D} \approx \varepsilon_d$, so that plasmons perceive a PMMA layer as a dielectric. On the other hand, at high enough frequencies at which $\varepsilon_d(\omega) > -\varepsilon_m(\omega)$ (this happens around $\lambda_0$~500 nm for a PMMA layer) $\varepsilon_{2D}$ changes sign and becomes negative. Thus, around $\lambda_0$~500 nm plasmons perceive a PMMA layer on gold as an “effective metal”. As a result, around $\lambda_0$~500 nm plasmons perceive a PMMA stripe pattern on gold substrate as a layered hyperbolic metamaterial shown in Fig.1(a). Fabrication of such plasmonic hyperbolic metamaterials in two dimensions requires only very simple and common lithographic techniques [4].

## 2. Super-resolution imaging using hyperbolic metamaterials: the hyperlens

Optical microscopy is one of the oldest research tools. Its development began in about 1590 with the observation by the Dutch spectacle maker Zaccharias Janssen and his son Hans that a combination of lenses in a tube made small objects appear larger. In 1609 Galileo Galilei improved on their ideas and developed an *occhiolino* or compound microscope with a convex and a concave lens The acknowledged “father” of microscopy is, however, Anton van Leeuwenhoek (1632-1723) who developed improved grinding and polishing techniques for making short focal length lenses, and was the first person to consequently see bacteria, protozoa, and blood cells.

Although various electron and scanning probe microscopes have long surpassed the compound optical microscope in resolving power, optical microscopy remains invaluable in many fields of science. The practical limit to the resolution of a conventional optical microscope is determined by diffraction: a wave cannot be localized to a region much smaller than half of its vacuum wavelength $\lambda_0/2$. Immersion microscopes introduced by Abbé in the 19th century have slightly improved resolution, on the order of $\lambda_0/2n$ because of the shorter wavelength of light $\lambda_0/n$ in a medium with refractive index $n$. However, immersion microscopes are limited by the small range of refractive indices $n$ of available transparent materials. For a while it was believed that the only way to achieve nanometer-scale spatial resolution in an optical microscope was to detect evanescent optical waves in very close proximity to a studied sample using a near-field scanning optical microscope (NSOM) [26]. Although many fascinating results are being obtained with NSOM, such microscopes are not as versatile and convenient to use as regular far-field optical microscopes. For example, an image from a near-field optical microscope is obtained by point-by-point scanning, which is an indirect and a rather slow process, and can be affected by artefacts of the sample.

An important early step to overcome this limitation was made in surface plasmon-assisted microscopy experiments [27], in which two-dimensional (2D) image magnification was achieved. In this microscope design the dispersion behavior of SPPs propagating in the boundary between a thin metal film and a dielectric (1.7) was exploited to use the 2-D optics of SPPs with a short wavelength to produce a magnified local image of an object on the surface. If the dispersion curve of SPPs on gold is examined, an excitation wavelength that provides a small group velocity gives rise to a 2-D SPP diffraction limit that is on the order of $\lambda_{SPP}=\lambda/n_g$. If a 2D “mirror” structure is fabricated in the surface, in the case of [27] using parabolic droplets on the surface (see Fig.4(b)), then SPPs propagating in the surface reflect or scatter at the boundaries of an object placed on the surface. These reflected or scattered SPPs are then imaged by the surface structure to produce a magnified 2-D image. This magnified image can be examined by a far-field microscope by using light scattered from surface roughness or from lithographically generated surface structures that scatter propagating SPPs in to far-field radiation. The increased spatial resolution of microscopy experiments performed with SPPs [27] is based on the “hyperbolic” dispersion law of such waves, which may be written in the form

$$k_{xy}^2 - |k_z|^2 = \frac{\varepsilon_d \omega^2}{c^2} \qquad (2.1)$$

where $\varepsilon_d$ is the dielectric constant of the medium bounding the metal surface, which for air is =1, $k_{xy}=k_p$ is the wave vector component in the plane of propagation, and $k_z$ is the wave vector component perpendicular to the plane. This form of the dispersion relation originates from the exponential decay of the surface wave field away from the propagation plane.

The "optical hyperlens" design described by Jacob et al. [2] extends this idea by using a hyperbolic metamaterial made of a concentric arrangement of metal and dielectric cylinders, which may be characterized by a strongly anisotropic dielectric permittivity tensor in which the tangential $\varepsilon_\theta$ and the radial $\varepsilon_r$ components have opposite signs. The resulting hyperbolic dispersion relation

$$\frac{k_r^2}{\varepsilon_\theta} - \frac{k_\theta^2}{|\varepsilon_r|} = \frac{\omega^2}{c^2} \tag{2.2}$$

does not exhibit any lower limit on the wavelength of propagating light at a given frequency. Therefore, in a manner similar to the 2D optics of SPPs, there is no usual diffraction limit in this metamaterial medium. Abbe's resolution limit simply does not exist. Optical energy propagates through such a metamaterial in the form of radial rays. Moreover, as demonstrated in Section 1, a pattern of polymethyl methacrylate (PMMA) stripes formed on a metal surface (as shown in Fig.4(a)) behaves as a 2D plasmonic equivalent of 3D hyperbolic metamaterial. Thus, two modes of operation of a 2D plasmonic microscope may be implemented, as shown in Figs. 4(a,b). A plasmon microscope may be operated in the "hyperlens mode" (Fig.4(a)) in which the plasmons generated by the sample located in the center of the plasmonic hyperlens propagate in the radial direction. The lateral distance between plasmonic rays grows with distance along the radius. The images are viewed by a regular microscope. Alternatively, a 2D plasmon microscope may be operated in the "geometrical optics" mode as shown in Fig.4(b). A nanohole array illuminated by external laser light may act as a source of surface plasmons, which are emitted in all directions. Upon interaction with the sample positioned near the focal point of the parabolically-shaped dielectric droplet, and reflection off the droplet edge, the plasmons form a magnified planar image of the sample. The image is viewed by a regular microscope. The droplet edge acts as an efficient plasmon mirror because of total internal reflection. It appears that both modes of operation exhibit strong evidence of optical superresolution.

The internal structure of the magnifying hyperlens (Fig.5(a)) consists of concentric rings of PMMA deposited on a gold film surface. The required concentric structures were defined using a Raith E-line electron beam lithography (EBL) system with ~70 nm spatial resolution. The written structures were subsequently developed using a 3:1 IPA/MIBK solution (Microchem) as developer and imaged using AFM (see Fig.4(a)). According to theoretical proposals in refs.[1,2], optical energy propagates through a hyperbolic metamaterial in the form of radial rays. This behaviour is clearly demonstrated in Fig.5(b). If point sources are located near the inner rim of the concentric metamaterial structure, the lateral separation of the rays radiated from these sources increases upon propagation towards the outer rim. Therefore, resolution of an "immersion" microscope (a hyperlens) based on such a metamaterial structure is defined by the ratio of inner to outer radii. Resolution appears limited only by losses, which can be compensated by optical gain.

The magnifying superlenses (or hyperlenses) have been independently realized for the first time in two experiments [4,5]. In particular, experimental data obtained using a 2D plasmonic hyperlens (shown in Fig.5(a)) do indeed demonstrate ray-like propagation of subwavelength plasmonic beams emanated by test samples. Far-field optical resolution of at least 70 nm (see Fig.6(f)) has been demonstrated using such a magnifying hyperlens based on a 2D plasmonic metamaterial design. Rows of either two or three PMMA dots have been produced near the inner ring of the hyperlens (Fig.6 b,c). These rows of PMMA dots had 0.5 μm periodicity in the radial direction so that phase matching between the incident laser light and surface plasmons can be achieved. Upon illumination with an external laser, the three rows of PMMA dots in Fig.6(b) gave rise to three divergent plasmon "rays", which are clearly visible in the plasmon image in Fig.6(d) obtained using a conventional optical microscope. The cross section analysis of this image across the plasmon "rays" (Fig.6(f)) indicates resolution of at least 70 nm or ~λ/7. The lateral separation between these rays increased by a factor of ten as the rays reached the outer rim of the hyperlens. This increase allowed visualization of the triplet using a conventional microscope. In a similar fashion, two rows of PMMA dots shown in Fig.6(c) gave rise to two plasmon rays, which are visualized in Fig.6(e).

The magnifying action and the imaging mechanism of the hyperlens have been further verified by control experiments presented in Fig. 7. The image shown in Fig,7(a) presents results of two actual imaging experiments (top portion of Fig. 7(a)) performed simultaneously with four control experiments seen at the bottom of the same image. In these experiments, two rows of PMMA dots have been produced near the inner ring of the hyperlens structures seen at the top and at the bottom of Fig. 7(a) (the AFM image of the dots is seen in the inset). These rows of PMMA dots had 0.5 μm periodicity in the radial direction so that phase matching between the incident 515 nm laser light and surface plasmons can be achieved. On the other hand, no such PMMA dot structure was fabricated near the control hyperlenses seen in the center of Fig. 7(a). Upon illumination with an external laser, the two rows of PMMA dots gave rise to the two divergent "plasmon rays," which are clearly visible in the top portion of the image in Fig.7(a) obtained using a conventional optical microscope. No such rays were observed in the four “control” hyperlenses visible in the bottom portion of the same image. There was no sample to image for the two hyperlenses located in the center of Fig. 7(a). On the other hand, the PMMA dot structure was designed for phase-matched plasmon generation in the “upward” direction as seen in the image. That is why no plasmon rays are visible when the hyperlens structures are inverted, as seen in the bottom of Fig. 7(a). When the gold film was replaced with an ITO film in another control experiment performed using the same experimental geometry, no hyperlens imaging occurred since no surface plasmons are generated on ITO surface (see Fig.7(b)). These experiments clearly verify the imaging mechanism and increased spatial resolution of the plasmonic hyperlens.

## 3. Consequences of singular photonic density of states: radiative decay engineering, thermal hyper-conductivity, and new stealth technologies

As we discussed in Section 1, the broadband divergence of photonic density of states in hyperbolic metamaterial is unique since it leads to the infinite value of the density of states for every frequency where different components of the dielectric permittivity have opposite signs. This very large number of electromagnetic states can couple to quantum emitters leading to such unusual phenomena as the broadband Purcell effect [6] and thermal hyperconductivity [10]. On the other hand, free space photons illuminating a roughened surface of hyperbolic metamaterial preferentially scatter inside the metamaterial leading to the surface being “darker than black” at the hyperbolic frequencies [9]. The latter property may find natural applications in stealth technologies.

As a first example of these unusual quantum behaviours, let us consider the broadband Purcell effect, which may become extremely useful for such applications as single photon sources, fluorescence imaging, biosensing and single molecule detection. In the spirit of the Fermi's golden rule, an increased number of radiative decay channels due to the high-k states in hyperbolic media (available for an excited atom) must ensure enhanced spontaneous emission. This enhancement can increase the quantum yield by overcoming emission into competing non-radiative decay routes such as phonons. A decrease in lifetime, high quantum yield and good collection efficiency can lead to extraction of single photons reliably at a high repetition rate from isolated emitters [28]. The available radiative channels for the spontaneous photon emission consist of the propagating waves in vacuum, the plasmon on the metamaterial substrate and the continuum of high wavevector waves which are evanescent in vacuum but propagating within the metamaterial. The corresponding decay rate into the metamaterial modes when the emitter is located at a distance $a<d<<\lambda$ (where $a$ is the metamaterial patterning scale) is [6]

$$\Gamma^{meta} \approx \frac{\mu^2}{2\hbar d^3} \frac{2\sqrt{\varepsilon_x |\varepsilon_z|}}{\left(1+\varepsilon_x |\varepsilon_z|\right)} \tag{3.1}$$

In the close vicinity of the hyperbolic metamaterial, the power from the dipole is completely concentrated in the large spatial wavevector channels (Fig 8(a) inset). The same evanescent wave spectrum when incident on a lossy metal or dielectric would be completely absorbed, causing a non-radiative decrease in the lifetime of an emitter (quenching). On the contrary, the metamaterial converts the evanescent waves to propagating and the absorption thus affects the outcoupling efficiency of the emitted photons due to a finite propagation length in the metamaterial.

Along with the reduction in lifetime and high efficiency of emission into the metamaterial, another key feature of the hyperbolic media is the directional nature of light propagation [6]. Fig. 8(b) shows the field along a plane perpendicular to the metamaterial-vacuum interface exhibiting the beamlike radiation from a point dipole. This is advantageous from the point of view of collection efficiency of light since the spontaneous emitted photons lie within a cone. The group velocity vectors in the medium which point in the direction of the Poynting vector are simply normals to the dispersion curve. For vacuum, these normals point in all directions and hence the spontaneous emission is isotropic in nature. In contrast to this behavior, the hyperbolic dispersion medium allows wave vectors only within a narrow region defined by the asymptotes of the hyperbola. Hence the group velocity vectors lie within the resonance cone giving rise to a directional spontaneously emitted photon propagating within the metamaterial.

Since its theoretical prediction in [6], the broadband Purcell effect has been indeed observed in multiple experiments, such as [7,8]. Virtually the same physics as in Fig.8(b) is also responsible for the "darker than black" behavior of roughened hyperbolic metamaterials [9]. Free space photons illuminating a rough surface of the hyperbolic metamaterial preferentially scatter inside the metamaterial into the bulk high k-vector modes. As a result, the photon probability to scatter back into free space is almost zero and the roughened surface looks black in the hyperbolic frequency bands.

Let us now consider radiative heat transfer inside hyperbolic metamaterials. It appears that the broadband divergence of the photonic density of states described above also leads to giant increase in radiative heat transfer compared to the Stefan-Boltzmann law in vacuum and in usual anisotropic dielectric materials. According to numerical calculations [10], this radiative thermal "hyperconductivity" may approach or even exceed heat conductivity via electrons and phonons in regular solids with the additional advantage of radiative heat transfer being much faster. Therefore, this radiative thermal hyperconductivity may potentially be very useful in fast microelectronics heat management applications [29]. In such applications heat generated by micro and nanoelectronic circuit components needs to be quickly dissipated at a heat sink, which cannot be located in the immediate vicinity of the electronic component. A hyperbolic metamaterial heat management layer may solve this important technological problem.

Let us start by tracing how the photonic density of states enters the usual Stefan-Boltzmann law. For the sake of simplicity, we will consider vacuum as a typical example of "normal" or "elliptical" material. As usual, we can start by calculating energy density of the black body radiation. A well-known textbook derivation can be summarized as follows:

$$u_{ell} = \frac{U}{V} = \int_0^\infty \frac{\varepsilon}{\exp\left(\frac{\varepsilon}{kT}\right) - 1} g(\varepsilon) d\varepsilon = \frac{4\sigma T^4}{c} \tag{3.2}$$

where $g(\varepsilon)$ is the photonic density of states. Eq.(3.2) clearly demonstrates that the drastic change in the density of states schematically shown in Fig.9 must lead to the drastic change in the final result. The singular behavior of the photonic density of states in hyperbolic metamaterial takes these media beyond the realm of the Stefan-Boltzmann law, with no ultimate limit on the radiative heat transfer. For the energy flux along the symmetry axis of a uniaxial hyperbolic metamaterial, it was found [10] that

$$S_T = \frac{\hbar c^2 k_{max}^4}{32\pi^2} \int d\omega \frac{1}{\exp\left(\frac{\hbar\omega}{kT}\right) - 1} \left| \frac{\varepsilon_1 \frac{d\varepsilon_2}{d\omega} - \varepsilon_2 \frac{d\varepsilon_1}{d\omega}}{\det\|\varepsilon\|} \right| \tag{3.3}$$

where the frequency integration is taken over the frequency bandwidth corresponding to the hyperbolic dispersion. Note that the heat flux in eqn. (3.3) is very sensitive to the dispersion in the hyperbolic metamaterial, $d\varepsilon/d\omega$. Indeed, the derivative of the dielectric permittivity determines the difference in the asymptotic behavior of the k-vector between the two hyperbolic surfaces that determine the phase space volume between the frequencies $\omega$ and $\omega + d\omega$ (see Fig. 9), and thus defines the actual value of the density of states. While there are many metamaterial designs leading to the hyperbolic dispersion, the most practical and widely used systems rely on either the medal-dielectric layer approach, or incorporate aligned metal nanowire composites (as shown in Fig.1). For the planar layers design, the hyperbolic behavior is observed for the wavelengths above ~10 μm if the system is fabricated using semiconductors [20], or for the wavelength above ~1 μm if the metamaterial is composed of metal-dielectric layers [7]. For the nanowire based approach, the hyperbolic dispersion is present at $\lambda > 1\mu m$ [21]. As a result, with either of these conventional metamaterial designs, the desired hyperbolic behavior covers the full range of wavelength relevant for the radiative heat transfer. As a result, the following estimates on the thermal energy flux in hyperbolic metamaterials have been obtained [10]:

$$S_T \approx \frac{\varepsilon_d}{4(1-n)} S_T^{(0)} \left( \frac{k_{\max}}{k_p} \right)^4 \tag{3.4}$$

for the layered material design, and

$$S_T \approx \frac{5}{16\pi^2} S_T^{(0)} \left( \frac{k^2{}_{\max}}{k_T k_p} \right)^2 \tag{3.5}$$

for the wire array design, where $S^{(0)}{}_T$ is the blackbody thermal energy flux for emission into the free space, $k_p$ is the plasma momentum, and $k_T$ is the thermal momentum. In both cases the numerical values of $S_T$ exceed $S^{(0)}{}_T$ by 4 to 5 orders of magnitude, thus firmly placing hyperbolic metamaterials in the realm of practical applications for radiative heat transfer and thermal management. A similar enhancement may be also expected in thermal conductivity.

## 4. Photonic hypercrystals

Explosive development of research on hyperbolic metamaterial also resulted in the recent demonstration of a novel artificial optical material, the "photonic hyper-crystal" [30], which combines the most interesting features of hyperbolic metamaterials and photonic crystals. Similar to hyperbolic metamaterials, photonic hyper-crystals exhibit broadband divergence in their photonic density of states due to the lack of usual diffraction limit on the photon wave vector. On the other hand, similar to photonic crystals, hyperbolic dispersion law of extraordinary photons is modulated by forbidden gaps near the boundaries of photonic Brillouin zones. Three dimensional self-assembly of photonic hyper-crystals has been achieved by application of external magnetic field to a cobalt nanoparticle-based ferrofluid. Unique spectral properties of photonic hyper-crystals lead to extreme sensitivity of the material to monolayer coatings of cobalt nanoparticles, which should find numerous applications in biological and chemical sensing.

Over the last few decades a considerable progress has been made in developing artificial optical materials with novel and often counterintuitive properties. Revolutionary research by Yablonovitch and John on photonic crystals [31,32] was followed by the development of electromagnetic metamaterial paradigm by Pendry [33]. Even though considerable difficulties still exist in fabrication of three-dimensional (3D) photonic crystals and metamaterials, both fields exhibit considerable experimental progress [34,35]. On the other hand, on the theoretical side these fields are believed to be complementary but mutually exclusive. Photonic crystal effects typically occur in artificial optical media which are periodically structured on the scale of free space light wavelength $\lambda$, while electromagnetic metamaterials are required to be structured (not necessarily in a periodic fashion) on the scale, which is much smaller than the free space wavelength of light. For example, in metal nanowire-based hyperbolic metamaterials schematically shown in Fig.1(b) the inter-wire distance must be much smaller than $\lambda$. Experimental realization of 3D "photonic hyper-crystals" [30] bridges this divide by combining the most interesting properties of hyperbolic metamaterials and photonic crystals.

The concept of the photonic hyper-crystal is based on the fact that dispersion law of extraordinary photons in hyperbolic metamaterials (1.2) does not exhibit the usual diffraction limit. Existence of large $k$-vector modes in a broad range of frequencies means that periodic modulation of hyperbolic metamaterial properties on a scale $L \ll \lambda$ (see inset in Fig.10(a)) would lead to Bragg scattering of extraordinary photons and formation of photonic band structure no matter how small $L$ is [30]. Thus, so formed "photonic hyper-crystals" would combine the most interesting properties of hyperbolic metamaterials and photonic crystals. For example, similar to classic photonic crystal effect predicted by John [32], strong localization of photons may occur in photonic hyper-crystals. However, unlike usual photonic crystals where light localization occurs on a scale $\sim\lambda$, photonic hyper-crystals may exhibit light localization on deep subwavelength scale. Similar to surface plasmon resonance (SPR) [25] and surface enhanced Raman (SERS) [36] - based sensing, engineered localization of light on deep subwavelength scale in photonic hyper-crystals should find numerous applications in biological and chemical sensing.

Band structure and field distribution inside a photonic hyper-crystal may be obtained in a straightforward manner. While both $\varepsilon_{xy}$ and $\varepsilon_z$ may exhibit periodic spatial dependencies, let us consider the relatively simple case of coordinate-independent $\varepsilon_{xy} > 0$ and periodic $\varepsilon_z(z) < 0$ with a period $L << \lambda$. Aside from the relative mathematical simplicity of this model, it also corresponds to the most readily available low-loss realizations of a hyperbolic metamaterials such as the composites formed by metallic nanowires in a dielectric membrane [21] (where $\varepsilon_{xy} > 0$ and $\varepsilon_z(z) < 0$), and planar layered metal-dielectric and semiconductor metamaterials [7,20]. Taking into account the translational symmetry of the system in $x$ and $y$ directions, we can introduce the in-plane wave vector ($k_x$, $k_y$) so that the propagating waves can be expressed as

$$E_\omega(\vec{r}) = E(z)\exp(ik_x x + ik_y y) \tag{4.1}$$

$$D_\omega(\vec{r}) = D(z)\exp(ik_x x + ik_y y)$$

$$B_\omega(\vec{r}) = B(z)\exp(ik_x x + ik_y y)$$

The uniaxial symmetry of this medium reduces the ordinary and extraordinary waves to respectively the TE ($\vec{E} \perp \hat{z}$) and TM ($\vec{B} \perp \hat{z}$) - polarized modes. Introducing the wavefunction $\psi(\vec{r})$ as the $z$-component of the electric displacement field of the TM wave

$$\psi(\vec{r}) = D_z(\vec{r}) = \varepsilon_z(z)E_z(\vec{r}) = -\frac{c}{\omega}k_x B \tag{4.2}$$

for the wave equation we obtain

$$-\frac{\partial^2\psi}{\partial z^2}+\frac{\varepsilon_{xy}}{\varepsilon_z(z)}\psi=\varepsilon_{xy}\frac{\omega^2}{c^2}\psi \tag{4.3}$$

In this wave equation the periodic $\varepsilon_{xy}/\varepsilon_z$ ratio acts as a periodic effective potential. As usual, solutions of eq.(4) may be found as Bloch waves

$$\psi(z)=\sum_{m=0}^{\infty}\psi_m \exp(i(k_z+\frac{2\pi}{L}m)z) \tag{4.4}$$

where $k_z$ is defined within the first Brillouin zone $-\pi/L<k_z<\pi/L$. Strong Bragg scattering is observed near the Brillouin zone boundaries at $k_z \sim \pi/L >> \pi/\lambda$, leading to the formation of photonic bandgaps in both the wavenumber and the frequency domains. This behavior is illustrated in Fig. 11, where we compare the dispersion diagram for an example of such nanowire-based photonic hyper-crystal to its effective medium counterpart. The material parameters of photonic hyper-crystal are based on the parameters of stratified ferrofluid described below. Similar to the usual photonic crystals [32], adiabatic chirping of *L* leads to strong field enhancement which, unlike that in the conventional photonic crystals, occurs on a deep subwavelength scale. We should also mention that the most interesting case appears to be the epsilon-near-zero (ENZ) situation where $\varepsilon_z$ approaches zero near a periodic set of planes. As has been demonstrated in Ref. [16], electric field of the extraordinary wave diverges in these regions. These periodic field divergences appear to be most beneficial for sensing applications.

Validation of the photonic hyper-crystal concept have been achieved using an experimental technique based on three-dimensional self-assembly of cobalt nanoparticles in the presence of external magnetic field [30]. Magnetic nanoparticles in a ferrofluid are known to form nanocolumns aligned along the magnetic field [37].

Moreover, depending on the magnitude of magnetic field, nanoparticle concentration and solvent used, phase separation into nanoparticle rich and nanoparticle poor phases may occur in many ferrofluids [38]. This phase separation occurs on a 0.1-1 micrometer scale. Therefore, it can be used to fabricate a self-assembled photonic hypercrystal.

These experiments used cobalt magnetic fluid 27-0001 from Strem Chemicals composed of 10 nm cobalt nanoparticles in kerosene coated with sodium dioctylsulfosuccinate and a monolayer of LP4 fatty acid condensation polymer. The average volume fraction of cobalt nanoparticles in this ferrofluid is p=8.2%. Cobalt behaves as an excellent metal in the long wavelength infrared range (LWIR), as evident by Fig. 12(a): the real part of its refractive index, $n$, is much smaller than its imaginary part, $k$ [39]. Thus, real part of $\varepsilon$, Re$\varepsilon = n^2 - k^2$, is negative, and its absolute value is much larger than its imaginary part, Im$\varepsilon = 2nk$. Therefore, it is highly suitable for fabrication of hyperbolic metamaterials. The structural parameter of such a metamaterial falls into a few nanometers range: the cobalt nanoparticle size is 10 nm, while average inter-particle distance at 8.2% volume fraction is about 19 nm. Therefore, the metamaterial properties may be described by effective medium parameters on spatial scales ~100 nm. On the other hand as demonstrated below, ferrofluid begins to exhibit hyperbolic behavior in the range of free space wavelengths ~10000 nm and above - in the so called long wavelength infrared (LWIR) frequency range. Thus, in between 100 nm and 10000 nm there exists an ample range of spatial scales which enable photonic hyper-crystal behavior described above. For example, if the effective medium parameters of ferrofluid are modulated on the scale of ~2000 nm (as evident from Fig.10(d)) the large k-vector modes which exist in the metamaterial in the LWIR range will experience Bragg scattering due to this modulation.

Electromagnetic properties of these metamaterials may be understood based on the Maxwell-Garnett approximation via the dielectric permittivities $\varepsilon_m$ and $\varepsilon_d$ of cobalt and kerosene, respectively, as illustrated in Fig.12. Volume fraction of cobalt nanoparticles aligned into nanocolumns by external magnetic field, α(B), depends on the field magnitude. At very large magnetic fields all nanoparticles are aligned into nanocolumns, so that $\alpha(\infty)=$p=8.2%. At smaller fields the difference $\alpha(\infty)$-α(B) describes cobalt nanoparticles, which are not aligned and distributed homogeneously inside the ferrofluid. Using this model, the diagonal components of the ferrofluid permittivity may be calculated and measured as a function of magnetic field. The value of $\alpha_H$ as a function of wavelength is plotted in Fig.3(b). This plot indicates that the original ferrofluid diluted with kerosene at a 1:10 ratio remains a hyperbolic medium above $\lambda$ = 5μm. More interestingly, such a diluted ferrofluid develops very pronounced phase separation into periodically aligned cobalt rich and cobalt poor phases (with periodicity L~2 μm) if subjected to external magnetic field. Optical microscope images of the diluted ferrofluid before and after application of external magnetic field are shown in Figs. 1(b) and 1(d). The periodic pattern of self-assembled stripes visible in image (d) appears due to phase separation. The stripes are oriented along the direction of magnetic field. The stripe periodicity $L$~2μm appears to be much smaller than the free space wavelength in the hyperbolic frequency range. Therefore, created self-assembled optical medium appears to be a photonic hyper-crystal. We should also note that the original undiluted ferrofluid exhibits similar phase separation in external magnetic field, even though on much smaller ~0.3 μm spatial scale (see Fig.1(f)).

Polarization dependencies of ferrofluid transmission as a function of magnetic field and nanoparticle concentration measured in a broad 0.5 μm - 16 μm wavelength range conclusively prove hyperbolic crystal character of ferrofluid anisotropy in the long wavelength IR range at large enough magnetic field. Fig. 3(e) shows polarization-dependent transmission spectra of 200 μm thick undiluted ferrofluid sample obtained using FTIR spectrometer. These data are consistent with hyperbolic character of $\varepsilon$ tensor of the ferrofluid in $B$ = 1000 G. Ferrofluid transmission is large for polarization direction perpendicular to magnetic field (perpendicular to cobalt nanoparticle chains) suggesting dielectric character of $\varepsilon$ in this direction. On the other hand, ferrofluid transmission falls to near zero for polarization direction along the chains, suggesting metallic character of $\varepsilon$ in this direction. However, these measurements are clearly affected by numerous ferrofluid absorption lines.

Fabricated photonic hyper-crystals exhibit all the typical features associated with the hyperbolic metamaterials. For example, absorption spectra measured using FTIR spectrometer with and without external magnetic field are consistent with the decrease of the radiation lifetime of kerosene molecules in the hyperbolic state. In addition, Fig. 13 clearly illustrates the photonic hyper-crystal potential in chemical and biological sensing, which is made possible by spatially selective field enhancement effects described above. This potential is revealed by detailed measurements of magnetic field induced transmission of photonic hyper-crystals in the broad IR spectral range presented in Figs. 13(a) and 13(b). FTIR spectral measurements are broadly accepted as a powerful “chemical fingerprinting” tool in chemical and biosensing. Therefore, broadly available magnetic field-tunable photonic hyper-crystals operating in the IR range open up new valuable opportunities in chemical analysis. The experimental data presented in Fig. 13 clearly illustrate this point. The FTIR transmission spectrum of the diluted (p=0.8%) ferrofluid in Fig.13(a) exhibits clear set of kerosene absorption lines, which is consistent with other published data (see for example ref.[40]). On the other hand, magnetic field induced transmission spectrum of the p=8.2% ferrofluid shown in Fig.13(b) contains a very pronounced absorption line at $\lambda$~12 $\mu$m (~840 $cm^{-1}$), which cannot be attributed to kerosene. Quite naturally this absorption line may be attributed to fatty acids, since cobalt nanoparticles are coated with a monolayer of surfactant composed of various fatty acids, such as lactic, oleic etc. acids as shown in Fig.13(d). Detailed comparison of Fig.13(b) with the 10-14 $\mu$m portion of lactic acid FTIR absorption spectrum shown in Fig.13(c) indeed indicates a close match. The fatty acid line appears to be about as strong as $\lambda$~14 $\mu$m (~695 $cm^{-1}$) line of kerosene, even though the oscillator strength of these molecular lines is about the same, while the amount of kerosene in the sample is ~ two orders of magnitude larger (a monolayer coating of fatty acids on a 10 nm cobalt nanoparticle occupies no more than 1% of ferrofluid volume). This paradoxical situation clearly indicates local field enhancement effects. Another strong evidence of field enhancement is provided by measurements of extinction coefficient of the ferrofluid presented in Fig.6E. Ferrofluid subjected to magnetic field exhibits pronounced resonances around the fatty acid absorption line at $\lambda$~12 $\mu$m and the kerosene absorption line at $\lambda$~14 $\mu$m. These resonances provide clear evidence of field enhancement by cobalt nanoparticle chains. We expect that further optimization of photonic hyper-crystals geometry will lead to much stronger sensitivity of their optical properties to chemical and biological inclusions, indicating a very strong potential of photonic hyper-crystals in biological and chemical sensing.

## 5. Superconducting hyperbolic metamaterials

Superconducting properties of a material, such as electron-electron interactions and the critical temperature of superconducting transition can be expressed via the effective dielectric response function $\varepsilon_{eff}(q,\omega)$ of the material. Such a description is valid on the spatial scales below the superconducting coherence length (the size of the Cooper pair), which equals ~100 nm in a typical BCS superconductor. Searching for natural materials exhibiting larger electron-electron interactions constitutes a traditional approach to high temperature superconductivity research. However, not long ago it was pointed out that the recently developed field of electromagnetic metamaterials deals with somewhat related task of dielectric response engineering on sub-100 nm scale, and that the metamaterial approach to dielectric response engineering may considerably increase the critical temperature of a composite superconductor-dielectric metamaterial [41]. Moreover, it appears that many high critical temperature (Tc) superconductors exhibit hyperbolic metamaterial properties in substantial portions of the electromagnetic spectrum of relevance to electon-electron interaction [42], so that their hyperbolicity may be partially responsible for their high Tc behavior.

Electromagnetic properties are known to play a very important role in the pairing mechanism and charge dynamics of high $T_c$ superconductors [43]. Moreover, shortly after the original work by Bardeen, Cooper and Schrieffer (BCS) [44], Kirzhnits *et al.* formulated a complementary description of superconductivity in terms of the dielectric response function of the superconductor [45]. The latter work was motivated by a simple argument that phonon-mitigated electron-electron interaction in superconductors may be expressed in the form of effective Coulomb potential

$$V(\vec{q},\omega)=\frac{4\pi e^2}{q^2\varepsilon_{eff}(\vec{q},\omega)}, \tag{5.1}$$

where $V=4\pi e^2/q^2$ is the usual Fourier-transformed Coulomb potential in vacuum, and $\varepsilon_{eff}(q,\omega)$ is the linear dielectric response function of the superconductor treated as an effective medium. Based on this approach, Kirzhnits et al. derived simple expressions for the superconducting gap $\Delta$, critical temperature $T_c$, and other important parameters of the superconductor. While thermodynamic stability condition implies [46] that $\varepsilon_{eff}(q,0)>0$, the dielectric response function at higher frequencies and spatial momenta is large and negative, which accounts for the weak net attraction and pairing of electrons in the superconducting condensate. In their paper Kirzhnits *et al.* noted that this effective medium consideration assumes "homogeneous system" so that "the influence of the lattice periodicity is taken into account only to the extent that it may be included into $\varepsilon_{eff}(q,\omega)$".

In the forty years which had passed since this very important remark, we have learned that the "homogeneous system" approximation may remain valid even if the basic structural elements of the material are not simple atoms or molecules. Now we know that artificial "metamaterials" may be created from much bigger building blocks, and the electromagnetic properties of these fundamental building blocks ("meta-atoms") may be engineered at will [47]. Since the superconducting coherence length (the size of the Cooper pair) is $\xi$~100 nm in a typical BCS superconductor, we have an opportunity to engineer the fundamental metamaterial building blocks in such a way that the effective electron-electron interaction (5.1) will be maximized, while homogeneous treatment of $\varepsilon_{eff}(q,\omega)$ will remain valid. In order to do this, the metamaterial unit size must fall within a rather large window between ~0.3 nm (given by the atomic scale) and $\xi$~100 nm scale of a typical Cooper pair. However, this task is much more challenging than typical applications of superconducting metamaterials suggested so far [48], which only deal with metamaterial engineering on the scales which are much smaller than the microwave or RF wavelength. Nevertheless, these experimental difficulties have been overcome in two recent successful demonstrations of metamaterial superconductors [49,50]. In the case of aluminium, its superconducting temperature was more than tripled by the metamaterial engineering [50].

Let us demonstrate that hyperbolic metamaterial geometry offers a natural way to increase attractive electron-electron interaction in a layered dielectric-superconductor metamaterial. Since hyperbolic metamaterials exhibit considerable dispersion, let us work in the frequency domain and write macroscopic Maxwell equations in the presence of "external" electron density $\rho_\omega$ and current $J_\omega$ as

$$\frac{\omega^2}{c^2}\vec{D}_\omega = \vec{\nabla}\times\vec{\nabla}\times\vec{E}_\omega - \frac{4\pi i\omega}{c^2}\vec{J}_\omega,\quad \vec{\nabla}\cdot\vec{D}_\omega = \rho_\omega,\quad \text{and}\quad \vec{D}_\omega = \vec{\varepsilon}_\omega \vec{E}_\omega \tag{5.2}$$

where the frequency $\omega$ is assumed to fall within the hyperbolic frequency band of the metamaterial. Let us solve eq.(5.2) for the z-component of electric field. After straightforward transformations we obtain

$$\frac{\omega^2}{c^2}E_z = \frac{4\pi}{\varepsilon_1\varepsilon_2}\frac{\partial\rho}{\partial z} - \frac{4\pi i\omega}{c^2\varepsilon_2}J_z - \frac{\partial^2 E_z}{\varepsilon_1\partial z^2} - \frac{1}{\varepsilon_2}\left(\frac{\partial^2 E_z}{\partial x^2} + \frac{\partial^2 E_z}{\partial y^2}\right) \tag{5.3}$$

Since $E_z = \partial\phi/\partial z$ , and the second term on the right side of eq.(5.3) may be neglected compared to the first one (since $v/c$<<1), we obtain

$$\frac{\omega^2}{c^2}\phi + \frac{\partial^2\phi}{\varepsilon_1\partial z^2} + \frac{1}{\varepsilon_2}\left(\frac{\partial^2\phi}{\partial x^2} + \frac{\partial^2\phi}{\partial y^2}\right) = \frac{4\pi}{\varepsilon_1\varepsilon_2}\rho \tag{5.4}$$

Taking into account that $V=-e\phi$, and neglecting the first term in eq(5.4) in the low frequency limit, we find that the effective Coulomb potential from eq.(5.1) assumes the form

$$V(\vec{q},\omega)=\frac{4\pi e^2}{q_z^2\varepsilon_2(\vec{q},\omega)+\left(q_x^2+q_y^2\right)\varepsilon_1(\vec{q},\omega)} \tag{5.5}$$

in a hyperbolic metamaterial. Since $\varepsilon_{xx}=\varepsilon_{yy}=\varepsilon_1$ and $\varepsilon_{zz}=\varepsilon_2$ have opposite signs, the effective Coulomb interaction of two electrons may become attractive and very strong in the hyperbolic frequency bands. The obvious condition for such a strong interaction to occur is

$$q_z^2\varepsilon_2(\vec{q},\omega)+\left(q_x^2+q_y^2\right)\varepsilon_1(\vec{q},\omega)\approx 0 \tag{5.6}$$

which indicates that the superconducting order parameter must be strongly anisotropic. This indeed appears to be the case in such hyperbolic high $T_c$ superconductors as BSCCO [42,43]. In order to be valid, the metamaterial "effective medium" description requires that the structural parameter of the metamaterial (in this particular case, the interlayer distance) must be much smaller than the superconducting coherence length. If the structural parameter approaches 1 nm scale, Josephson tunneling across the dielectric layers will become very prominent in such an anisotropic layered superconducting hyperbolic metamaterial.

The diagonal dielectric permittivity components of the layered superconductor-dielectric metamaterial may be calculated using Maxwell-Garnett approximation using eq.(1.5). In order to obtain hyperbolic properties, $\varepsilon_d$ of the dielectric needs to be very large, since $\varepsilon_m$ of the superconducting component given by the Drude model is negative and very large in the far infrared and THz ranges.

$$\varepsilon_m=\varepsilon_{m\infty}-\frac{\omega_p^2}{\omega^2}\approx-\frac{\omega_p^2}{\omega^2} \tag{5.7}$$

This is consistent with the measured dielectric behavior of the parent BSCCO perovskite compound [43]. Moreover, if the high frequency behavior of $\varepsilon_d$ may be assumed to follow the Debye model [51]:

$$\mathrm{Re}\,\varepsilon_d = \frac{\varepsilon_d(0)}{1+\omega^2\tau^2} \approx \frac{\varepsilon_d(0)}{\omega^2\tau^2}, \tag{5.8}$$

broadband hyperbolic behavior arise due to similar $\sim\omega^{-2}$ functional behavior of $\varepsilon_d$ and $\varepsilon_m$ in the THz range.

As follows from eq.(1.5), if the volume fraction of metallic phase $n$ is kept constant, the hyperbolic behavior may occur only within the following range of plasma frequency $\omega_p^2$of the metallic phase:

$$\frac{\omega_p^2\tau^2}{\varepsilon_d(0)} \in \left[\frac{n}{1-n};\frac{1-n}{n}\right] \quad , \tag{5.9}$$

Otherwise, either $\varepsilon_1$ and $\varepsilon_2$ will be both positive if $\omega_p^2$ is too small, or both negative if $\omega_p^2$ is too large. Interestingly enough, the boundaries of superconducting and hyperbolic states in high Tc cuprates seem to overlap. Moreover, the crystallographic lattice of BSCCO shown in Fig.14 looks very similar to the geometry of a multilayer hyperbolic metamaterial. Indeed, in BSCCO the anisotropy of DC conductivity may reach $10^4$ for the ratio of in plane to out of plane conductivity in high quality single crystal samples. Polarization-dependent AC reflectance spectra measured in the THz and far-infrared frequency ranges [43] also indicate extreme anisotropy. In the normal state of high Tc superconductors the in-plane AC conductivity exhibits Drude-like behavior with a plasma edge close to 10000cm$^{-1}$, while AC conductivity perpendicular to the copper oxide planes is nearly insulating. Extreme anisotropy is also observed in the superconducting state. The typical values of measured in-plane and out of plane condensate plasma frequencies in high Tc superconductors are $\omega_{p,ab}$=4000-10000cm$^{-1}$, and $\omega_{p,c}$=1-1000cm$^{-1}$, respectively [43]. The measured anisotropy is the strongest in the BSCCO superconductors. These experimental measurements strongly support the qualitative picture of BSCCO structure as a layered hyperbolic metamaterial (Fig.14(b)) in which the copper oxide layers may be represented as metallic layers, while the SrO and BiO layers may be represented as the layers of dielectric. Based on these measured material parameters, the diagonal components of BSCCO dielectric tensor may be calculated [41,42] as shown in Fig.15. The appearance of hyperbolic bands is quite generic in high Tc superconductors. While examples of such natural hyperbolic high $T_c$ superconductors appear to fit well into the metamaterial scheme described above, it would be interesting to try and follow the metamaterial recipe in making novel "designer" superconductors.

## 6. Gravitation theory analogs based on hyperbolic metamaterials

Modern developments in gravitation research strongly indicate that classic general relativity is an effective macroscopic field theory, which needs to be replaced with a more fundamental theory based on yet unknown microscopic degrees of freedom. On the other hand, our ability to obtain experimental insights into the future fundamental theory is strongly limited by low energy scales available to terrestrial particle physics and astronomical observations. The emergent analogue spacetime program offers a promising way around this difficulty. Looking at such systems as superfluid helium and cold atomic Bose-Einstein condensates, physicists learn from Nature and discover how macroscopic field theories arise from known well-studied atomic degrees of freedom. Another exciting development along this direction is recent introduction of metamaterials and transformation optics. The latter field is not limited by the properties of atoms and molecules given to us by Nature. "Artificial atoms" used as building blocks in metamaterial design offer much more freedom in constructing analogues of various exotic spacetime metrics, such as black holes [52-56], wormholes [57,58], spinning cosmic strings [59], and even the metric of Big Bang itself [15]. Explosive development of this field promises new insights into the fabric of spacetime, which cannot be gleaned from any other terrestrial experiments.

On the other hand, compared to standard general relativity, metamaterial optics gives more freedom to design an effective space-time with very unusual properties. Light propagation in all static general relativity situations can be mimicked with positive $\varepsilon_{ik} = \mu_{ik}$ [60], while the allowed parameter space of the metamaterial optics is broader. Thus, flat Minkowski space-time with the usual (-,+,+,+) signature does not need to be a starting point. Other effective signatures, such as the “two times” (2T) physics (-,-,+,+) signature may be realized [3]. Theoretical investigation of the 2T higher dimensional space-time models had been pioneered by Paul Dirac [61]. More recent examples can be found in [62,63]. Metric signature change events (in which a phase transition occurs between say (-,+,+,+) and (-,-,+,+) space-time signature) are being studied in Bose-Einstein condensates and in some modified gravitation theories (see ref.[64], and the references therein). It is predicted that a quantum field theory residing on a spacetime undergoing a signature change reacts violently to the imposition of the signature change. Both the total number and the total energy of the particles generated in a signature change event are formally infinite [64]. While optics of bulk hyperbolic metamaterials provides us with ample opportunities to observe metric signature transitions [3], even more interesting physics arise at the metamaterial interfaces. Very recently it was demonstrated that mapping of monochromatic extraordinary light distribution in a hyperbolic metamaterial along some spatial direction may model the “flow of time” in a three dimensional (2+1) effective Minkowski spacetime [15]. If an interface between two metamaterials is engineered so that the effective metric changes signature across the interface, two possibilities may arise. If the interface is perpendicular to the time-like direction $z$, this coordinate does not behave as a “timelike” variable any more, and the continuous “flow of time” is interrupted. This situation (which cannot be realized in classic general relativity) may be called the “end of time”. It appears that optics of metamaterials near the “end of time” event is quite interesting and deserves a detailed study. For example, in the lossless approximation all the possible “end of time” scenarios lead to field divergencies, which indicate quite interesting linear and nonlinear optics behaviour near the “end of time”. On the other hand, if the metamaterial interface is perpendicular to the space-like direction of the effective (2+1) Minkowski spacetime, a Rindler horizon may be observed (Rindler metric approximates spacetime behaviour near the black hole event horizon [60]).

Let us briefly summarize refs.[3,15], which demonstrated that a spatial coordinate may become “timelike” in a hyperbolic metamaterial. To better understand this effect, let us start with a non-magnetic uniaxial anisotropic material with dielectric permittivities $\varepsilon_x = \varepsilon_y = \varepsilon_1$ and $\varepsilon_z = \varepsilon_2$, and assume that this behaviour holds in some frequency range around $\omega=\omega_0$. Let us consider the case of constant $\varepsilon_1 > 0$ and $\varepsilon_2 < 0$, and assume that the metamaterial is illuminated by coherent CW laser field at frequency $\omega_0$. We will study spatial distribution of the extraordinary field $\varphi_\omega$ at this frequency. Under these assumptions equation (1.1) may be re-written in the form of 3D Klein-Gordon equation describing a massive scalar $\varphi_\omega$ field:

$$-\frac{\partial^2\varphi_\omega}{\varepsilon_1\partial z^2}+\frac{1}{|\varepsilon_2|}\left(\frac{\partial^2\varphi_\omega}{\partial x^2}+\frac{\partial^2\varphi_\omega}{\partial y^2}\right)=\frac{\omega_0{}^2}{c^2}\varphi_\omega=\frac{m^{*2}c^2}{\hbar^2}\varphi_\omega \tag{6.1}$$

in which the spatial coordinate $z=\tau$ behaves as a "timelike" variable. Therefore, eq.(6.1) describes world lines of massive particles which propagate in a flat (2+1) Minkowski spacetime. When a metamaterial is built and illuminated with a coherent extraordinary CW laser beam, the stationary pattern of light propagation inside the metamaterial represents a complete "history" of a toy (2+1) dimensional spacetime populated with particles of mass $m^*$. This "history" is written as a collection of particle world lines along the "timelike" $z$ coordinate. Note that in the opposite situation in which $\varepsilon_1$ <0 and $\varepsilon_2$ >0, equation (6.1) would describe world lines of tachyons [65] having "imaginary" mass $m^*=i\mu$.. Eq.(6.1) exhibits effective Lorentz invariance under the coordinate transformation

$$z'=\frac{1}{\sqrt{1-\frac{\varepsilon_{xy}}{(-\varepsilon_z)}\beta}}(z-\beta x) \tag{6.2}$$

$$x'=\frac{1}{\sqrt{1-\frac{\varepsilon_{xy}}{(-\varepsilon_z)}\beta}}\left(x-\beta\frac{\varepsilon_{xy}}{(-\varepsilon_z)}z\right),$$

where $\beta$ is the effective boost. Similar to our own Minkowski spacetime, the effective Lorentz transformations in the $xz$ and $yz$ planes form the Poincare group together with translations along $x$, $y$, and $z$ axis, and rotations in the $xy$ plane.

The world lines of particles described by eq.(6.1) are straight lines, which is easy to observe in the experiment [15]. If adiabatic variations of $\varepsilon_1$ and $\varepsilon_2$ are allowed inside the metamaterial, world lines of massive particles in some well-known curvilinear spacetimes can be emulated, including the world line behavior near the "beginning of time" at the moment of Big Bang, as illustrated in Fig.16 [15]. Thus, mapping of monochromatic extraordinary light distribution in a hyperbolic metamaterial along some spatial direction may model the "flow of time" in an effective three dimensional (2+1) spacetime. Since the parameter space of metamaterial optics is broader than the parameter space of general relativity, we can also engineer the "end of time" event if an interface between two metamaterials is prepared so that the effective metric changes signature at the interface. In such a case the spatial coordinate does not behave as a "timelike" variable any more, and the continuous "flow of time" is suddenly interrupted, as shown in Fig.17(a). This situation (which cannot be realized in classic general relativity) may be called the "end of time". It appears that optics of metamaterials near the "end of time" event is quite interesting and deserves a detailed study [16]. It appears that the optical images of field distribution over the sample surface indicate considerable field enhancement near the presumed plasmonic "end of time" events, as indicated by an arrow in Fig.18(f). On the other hand, a hyperbolic metamaterial interface, which is oriented perpendicular to the "space-like" direction (Fig.17(b)) behaves as a Rindler event horizon.

It is also interesting to note that nonlinear light propagation through a hyperbolic metamaterial may be formulated in a similar fashion as general relativity [13]. Sub-wavelength confinement of light in nonlinear hyperbolic metamaterials due to formation of spatial solitons has attracted much recent attention because of its seemingly counter-intuitive behaviour [66,67]. In order to achieve self-focusing in a hyperbolic wire medium, a nonlinear self-defocusing Kerr medium must be used as a dielectric host. Ref.[13] demonstrated that this behaviour finds natural explanation in terms of analogue gravity. Since wave equation describing propagation of extraordinary light inside hyperbolic metamaterials exhibits 2+1 dimensional Lorentz symmetry, we may assume that nonlinear optical Kerr effect “bends” this spacetime resulting in effective gravitational force between extraordinary photons. In order for the effective gravitational constant to be positive, negative self-defocusing Kerr medium must be used as a host. If gravitational self-interaction is strong enough, spatial soliton may collapse into a black hole analogue.

When the nonlinear optical effects become important, they are described in terms of various order nonlinear susceptibilities $\chi^{(n)}$ of the metamaterial:

$$D_i = \chi_{ij}^{(1)} E_j + \chi_{ijl}^{(2)} E_j E_l + \chi_{ijlm}^{(3)} E_j E_l E_m + ... \tag{6.3}$$

Taking into account these nonlinear terms, the dielectric tensor of the metamaterial (which defines its effective metric) may be written as

$$\varepsilon_{ij} = \chi_{ij}^{(1)} + \chi_{ijl}^{(2)} E_l + \chi_{ijlm}^{(3)} E_l E_m + ... \tag{6.4}$$

It is clear that eq.(6.4) provides coupling between the matter content (photons) and the effective metric of the metamaterial “spacetime”. However, in order to emulate gravity, the nonlinear susceptibilities $\chi^{(n)}$ of the metamaterial need to be engineered in some particular way. In the weak gravitational field limit the Einstein equation

$$R_i^k = \frac{8\pi\gamma}{c^4}\left(T_i^k - \frac{1}{2}\delta_i^k T\right) \tag{6.5}$$

is reduced to

$$R_{00} = \frac{1}{c^2}\Delta\phi = \frac{1}{2}\Delta g_{00} = \frac{8\pi\gamma}{c^4} T_{00} \quad , \tag{6.6}$$

where $\phi$ is the gravitational potential [60]. Since in our effective Minkowski spacetime $g_{00}$ is identified with $-\varepsilon_1$, comparison of eqs. (6.4) and (6.5) indicates that all the second order nonlinear susceptibilities $\chi^{(2)}_{ijl}$ of the metamaterial must be equal to zero, while the third order terms may provide correct coupling between the effective metric and the energy-momentum tensor. These terms are associated with the optical Kerr effect. All the higher order $\chi^{(n)}$ terms must be zero at $n>3$. Indeed, detailed analysis indicates [13] that Kerr effect in a hyperbolic metamaterial leads to effective gravity if the dielectric medium used as a metamaterial host exhibits self-defocusing nonlinearity, so that the magnetic ferrofluids described above appear to be ideal candidates to exhibit such phenomena.

Unlike other typical metamaterial systems, such ferrofluid-based macroscopic self-assembled 3D metamaterials may also exhibit reach physics associated with topological defects [14,68] and phase transitions. Therefore, as was pointed out recently by Mielczarek and Bojowald [62,63], the properties of self-assembled magnetic nanoparticle-based hyperbolic metamaterials exhibit strong similarities with the properties of some microscopic quantum gravity models, such as loop quantum cosmology. As described in Section 4, in the presence of an external magnetic field the ferrofluid forms a self-assembled hyperbolic metamaterial, which may be described as an effective "3D Minkowski spacetime" for extraordinary photons. If the magnetic field is not strong enough, this effective Minkowski spacetime gradually melts under the influence of thermal fluctuations, as illustrated in Fig.19. Thus, unlike other systems exhibiting analogue gravity behaviour, the unique feature of the ferrofluid consists in our ability to directly visualize the effective Minkowski spacetime formation at the microscopic level.

## 7. Summary

The diverse physical properties and applications of hyperbolic metamaterials outlined above clearly demonstrate that this sub-field of electromagnetic metamaterials already extended far beyond its original goal to enable sub-diffraction superresolution imaging. Hyperbolic metamaterials demonstrate a large number of novel phenomena resulting from the broadband singular behavior of their density of photonic states, which encompass enhanced quantum-electrodynamic effects, new stealth technology, thermal hyperconductivity, high Tc superconductivity, and very interesting gravitation theory analogues. Moreover, hyperbolic metamaterial behavior appears to be compatible with photonic crystal effects, such as deep sub-wavelength light localization and giant field enhancement, resulting in a fascinating new class of artificial optical media – the photonic hypercrystals. Since all this progress has been demonstrated within a few years from their inception, the future of hyperbolic metamaterials and their applications look really bright.

**Figure Captions**

**Figure 1.** Typical geometries of hyperbolic metamaterials: (a) multilayer metal-dielectric structure, and (b) metal wire array structure.

**Figure 2.** The constant frequency surfaces for (a) isotropic dielectric ($\varepsilon_1 = \varepsilon_2 > 0$) and (b) uniaxial hyperbolic ($\varepsilon_x = \varepsilon_y = \varepsilon_1 > 0$, $\varepsilon_z = \varepsilon_2 < 0$) metamaterial. (c) The phase space volume between two constant frequency surfaces for the hyperbolic metamaterial.

**Figure 3**. Dispersion law of surface plasmon polaritons in the lossless approximation.

**Figure 4**. Two modes of operation of a 2D plasmonic microscope. (a) Plasmon microscope operating in the "hyperlens mode": the plasmons generated by the sample located in the center of the hyperlens propagate in the radial direction. The lateral distance between plasmonic rays grows with distance along the radius. The images are viewed by a regular microscope. (b) Plasmon microscope operating in the "geometrical optics" mode: Nanohole array illuminated by external laser light acts as a source of surface plasmons, which are emitted in all directions. Upon interaction with the sample positioned near the focal point of the parabolically-shaped dielectric droplet, and reflection off the droplet edge, the plasmons form a magnified planar image of the sample. The image is viewed by a regular microscope. The droplet edge acts as an efficient plasmon mirror because of total internal reflection.

**Figure 5.** (a) Superposition image composed of an AFM image of the PMMA on gold plasmonic metamaterial structure superimposed onto the corresponding optical image obtained using a conventional optical microscope illustrating the imaging mechanism of the magnifying hyperlens. Near the edge of the hyperlens the separation of three rays (marked by arrows) is large enough to be resolved using a conventional optical

microscope. (b) Theoretical simulation of ray propagation in the magnifying hyperlens microscope.

**Figure 6.** AFM (a-c) and conventional optical microscope (d,e) images of the resolution test samples composed of three (a,b) and two (c) rows of PMMA dots positioned near the center of the magnifying hyperlens. The conventional microscope images presented in (d) and (e) correspond to the samples shown in (b) and (c), respectively. The rows of PMMA dots give rise to either three or two divergent plasmon "rays", which are visible in the conventional optical microscope images. (f) The cross section of the optical image along the line shown in (d) indicates resolution of at least 70 nm or ~ $\lambda/7$.

**Figure 7**. (a) This image obtained using a conventional optical microscope presents the results of two imaging experiments (top portion of the image) performed simultaneously with four control experiments seen at the bottom of the same image. The rows of PMMA dots shown in the inset AFM image were fabricated near the two top and two bottom hyperlenses. No such pattern was made near the two hyperlenses visible in the center of the image. Upon illumination with an external laser, the two rows of PMMA dots separated by 130 nm gap gave rise to two divergent plasmon rays shown by the arrows, which are clearly visible in the top portion of the image. The four control hyperlenses visible at the bottom do not produce such rays because there is no sample to image for the two hyperlenses in the center, and the two bottom hyperlenses are inverted. (b) Same pattern produced on ITO instead of gold film demonstrates a pattern of ordinary light scattering by the structure without any hyperlens imaging effects.

**Figure 8**. (a) Spontaneous emission lifetime of a perpendicular dipole above a hyperbolic metamaterial substrate (see inset). Note the lifetime goes to zero in the close vicinity of the metamaterial as the photons are emitted nearly instantly. Most of the power emitted by the dipole is concentrated in the large spatial modes (evanescent in vacuum) which are converted to propagating waves within the metamaterial. (inset) (b) False color plot of the field of the point dipole in a plane perpendicular to the metamaterial-vacuum interface (see inset of (a)) depicting the highly directional nature of the spontaneous emission (resonance cone).

**Figure 9**. The phase space volume between two constant frequency surfaces for (a) dielectric (elliptical) and (b) hyperbolic material with $\varepsilon_2>0$ and $\varepsilon_1<0$ (cut-out view). Panels (c) and (d) schematically illustrate different thermal conductivity mechanisms in (c) regular media (metals and dielectric) and (d) hyperbolic media. Giant radiative contribution to thermal conductivity in hyperbolic media can dominate the thermal transport.

**Figure 10.** (a) Experimental geometry of the ferrofluid-based hyperbolic metamaterial. The array of self-assembled cobalt nanocolumns has typical separation $a$~20 nm between the nanocolumns. The inset shows a photonic hyper-crystal structure formed by periodic arrangement of cobalt rich and cobalt sparse regions with typical periodicity $L$~2 μm, so that periodic modulation of hyperbolic metamaterial properties on a scale $L<<\lambda$ is achieved in the LWIR spectral range where $\lambda$~10 μm. Since photon wave vector in hyperbolic metamaterials is not diffraction-limited, periodic modulation of hyperbolic metamaterial properties on a scale $L<<\lambda$ would lead to Bragg scattering and formation of band structure. (b-f) Microscopic images of cobalt nanoparticle-based ferrofluid reveal subwavelength modulation of its spatial properties: frames (b) and (d) show microscopic images of the diluted cobalt nanoparticle-based ferrofluid before and after application of external magnetic field. The pattern of self-assembled stripes visible in image (d) is due to phase separation of the ferrofluid into cobalt reach and cobalt poor

phases. The stripes are oriented along the direction of magnetic field. The inset shows Fourier transform image of frame (d). Its cross section presented in panel (e) shows a histogram of different periods present in the image. A microscopic image of the sample taken along the axes of the nanowires is shown in frame (c). Panel (f) demonstrates that the original undiluted ferrofluid exhibits similar phase separation in external magnetic field, even though on much smaller scale.

**Figure 11.** Comparison of the effective medium dispersion of a nanowire-based hyperbolic metamaterial (a) to the exact solution for photonic crystal (b). The hyper-crystal unit cell in (b) is assumed to be 1000 nm, with one half of the cell is filled with the same ferrofluid while another half is pure kerosene. The ferrofluid aligned by external magnetic field applied in the z-direction, is characterized by the material parameters described in Fig. 12. The stratified ferrofluid is assumed to form layers with the normal along the x-direction.  The effective medium dispersion in panel (a) uses the dielectric permittivity tensor obtained by the homegenization of the electromagnetic response of the hyper-crystal unit cell. This is an artificial result that would be expected if we could experimentally isolate magnetic field induced hyperbolic behaviour of the ferrofluid from its photonic hyper-crystal behaviour. In both panels, the wavevector components are given in units of the free-space wavenumber $k_0$ .

**Figure 12.** (a) Optical properties of cobalt as tabulated in Ref. [22]: real (n) and imaginary (k) parts of the cobalt refractive index are plotted in the long wavelength IR range. (b) Critical volume fraction of cobalt nanoparticles corresponding to ferrofluid transition to hyperbolic metamaterial phase shown as a function of free space light wavelength. (c,d) Wavelength dependencies of $\varepsilon_z$ and $\varepsilon_{xy}$ at $\alpha(\infty)$ =8.2%. While $\varepsilon_{xy}$ stays positive and almost constant, $\varepsilon_z$ changes sign around $\lambda$=3μm. (e) Polarization-

dependent transmission spectra of 200 μm - thick ferrofluid sample measured using FTIR spectrometer are consistent with hyperbolic character of ε tensor.

**Figure 13.** (a) FTIR transmission spectrum of diluted ($\alpha(\infty)$=0.8%) ferrofluid exhibits clear set of kerosene absorption lines. (b) Transmission spectra of the $\alpha(\infty)$=8.2% ferrofluid measured with and without application of external magnetic field. Magnetic field induced transmission spectrum contains a very pronounced absorption line at λ~12 μm (~840 $cm^{-1}$), which can be attributed to lactic acid. Kerosene absorption lines are marked with yellow boxes, while the fatty acid line at 840 $cm^{-1}$ is marked with a green box. (c) The 10-14 μm portion of lactic acid FTIR absorption spectrum. (d) Schematic view of cobalt nanoparticle coated with a monolayer of fatty acids, such as lactic and oleic acid. (e) Extinction coefficient (the ratio T(0)/T(90) of ferrofluid transmissions at 0 and 90 degrees) of ferrofluid subjected to external magnetic field exhibits pronounced resonances around the fatty acid absorption line at λ~12 μm and the kerosene absorption line at λ~14 μm. These resonances provide clear evidence of field enhancement by cobalt nanoparticle chains.

**Figure 14.** Comparison of the crystallographic unit cell of a BSCCO high $T_c$ superconductor (a) and geometry of a layered hyperbolic metamaterial (b).

**Figure 15**. Diagonal components of the permittivity tensor of a high Tc BSCCO superconductor calculated as a function of frequency. The hyperbolic bands appear at $200cm^{-1}<\omega<1200cm^{-1}$ and $2400cm^{-1}<\omega<4800cm^{-1}$.

**Figure 16**. Experimental demonstration of world line behavior in an "expanding universe" using a plasmonic hyperbolic metamaterial: Optical (a) and AFM (b) images of the plasmonic hyperbolic metamaterial based on PMMA stripes on gold. The defect used as a plasmon source is shown by an arrow. (c) Plasmonic rays or "world lines"

increase their spatial separation as a function of "timelike" radial coordinate. The point (or moment) $r=\tau=0$ corresponds to a toy "big bang". For the sake of clarity, light scattering by the edges of the PMMA pattern is partially blocked by semi-transparent triangles. (d) Schematic view of world lines behavior near the Big Bang.

**Figure 17.** (a) Schematic representation of the "end of time" model in metamaterials: the spatial coordinate $z$ does not behave as a "timelike" variable any more, and the continuous "flow of time" is suddenly interrupted at the interface of two metamaterials. (b) Metric signature change across a "spacelike" direction leads to appearance of a Rindler horizon.

**Figure 18.** Experimental observation of the "end of time event" in a plasmonic hyperbolic metamaterial illuminated with 488 nm light.

**Figure 19.** Magnified images of the ferrofluid taken as a function of external magnetic field reveal fine details of the effective Minkowski spacetime melting. Top row shows these images in the usual greyscale format, while the quasi-3D representation of the same images in the bottom row provides better visualization of the actual nanoparticle filaments.

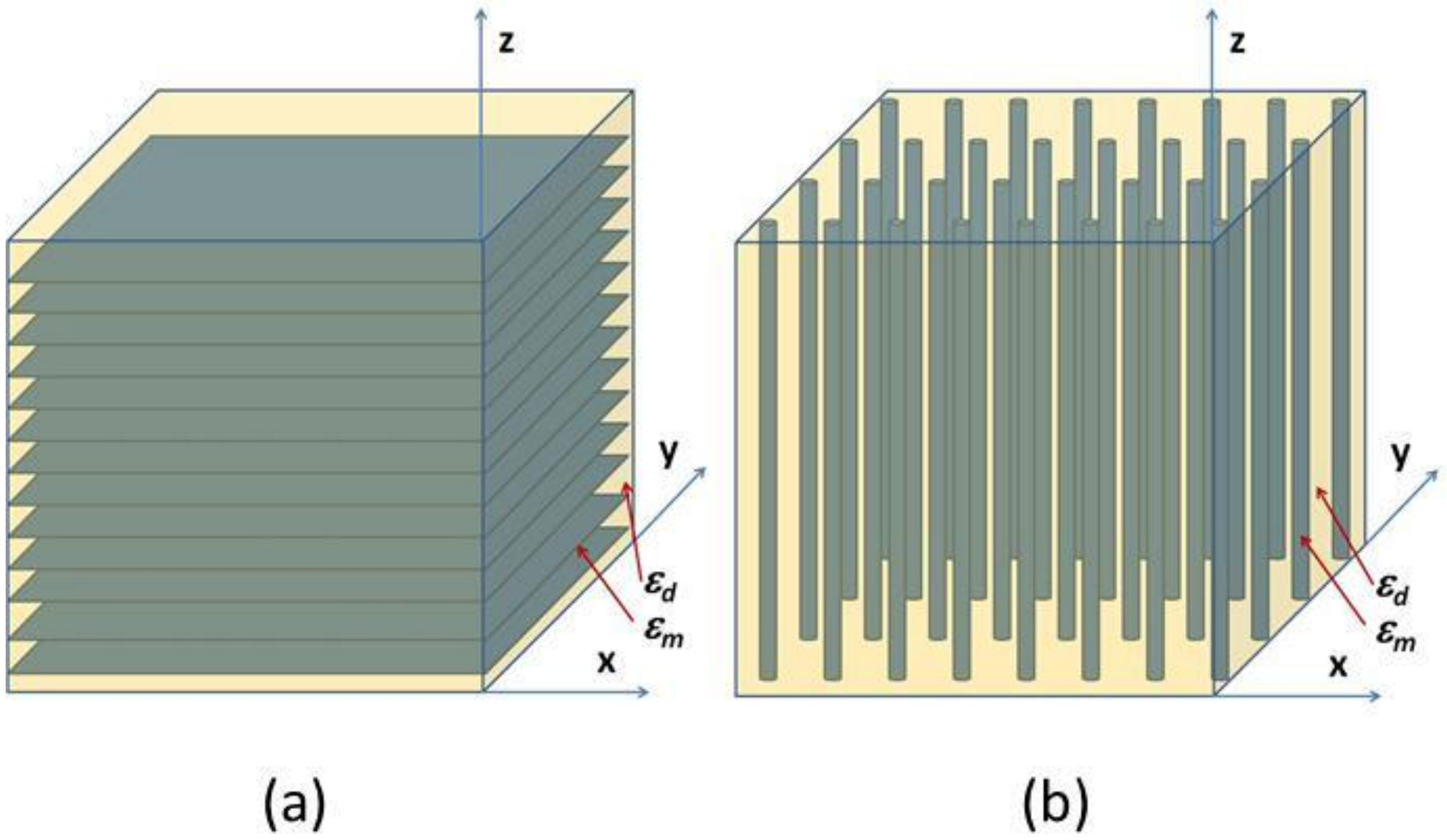

Fig. 1

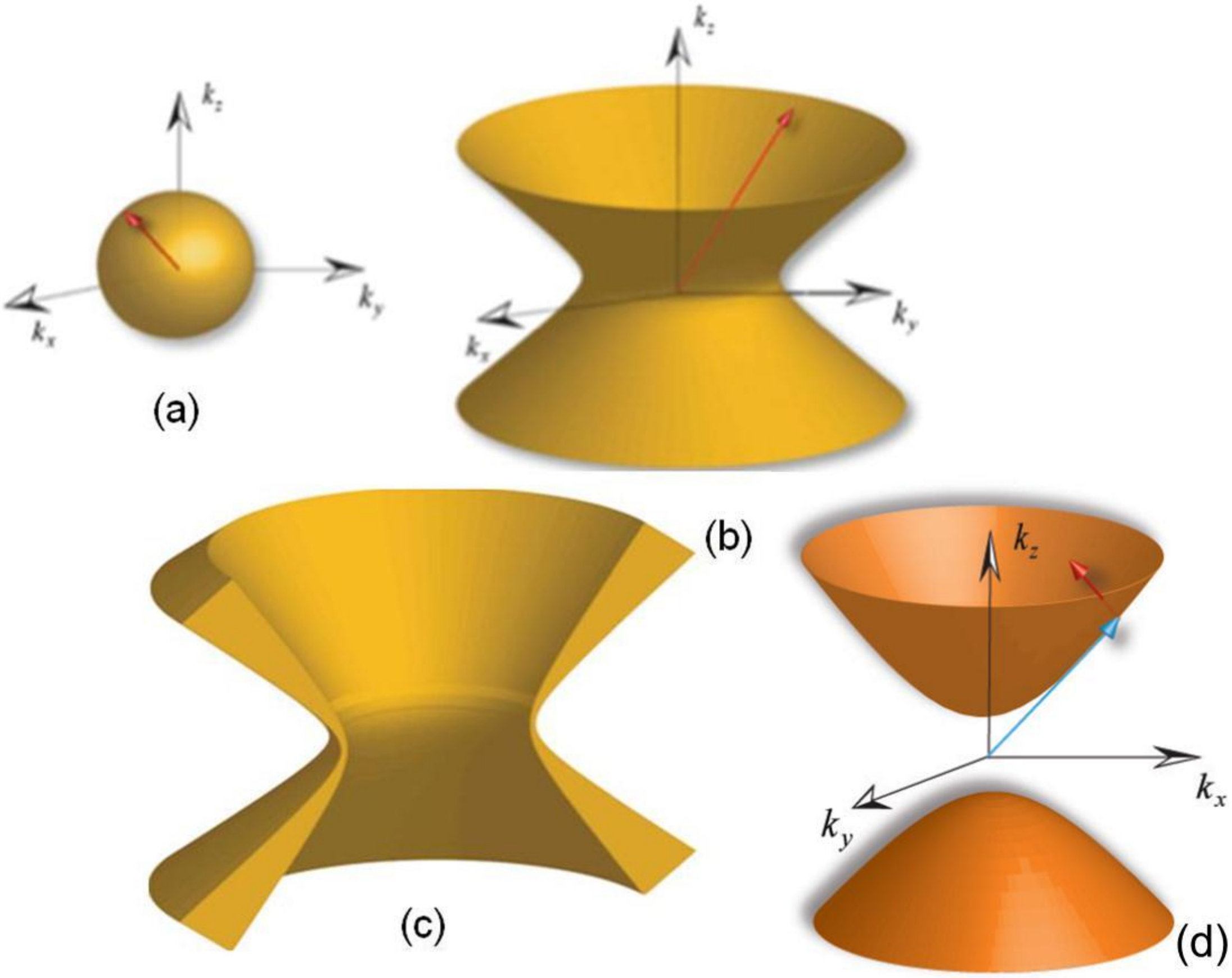

Fig. 2

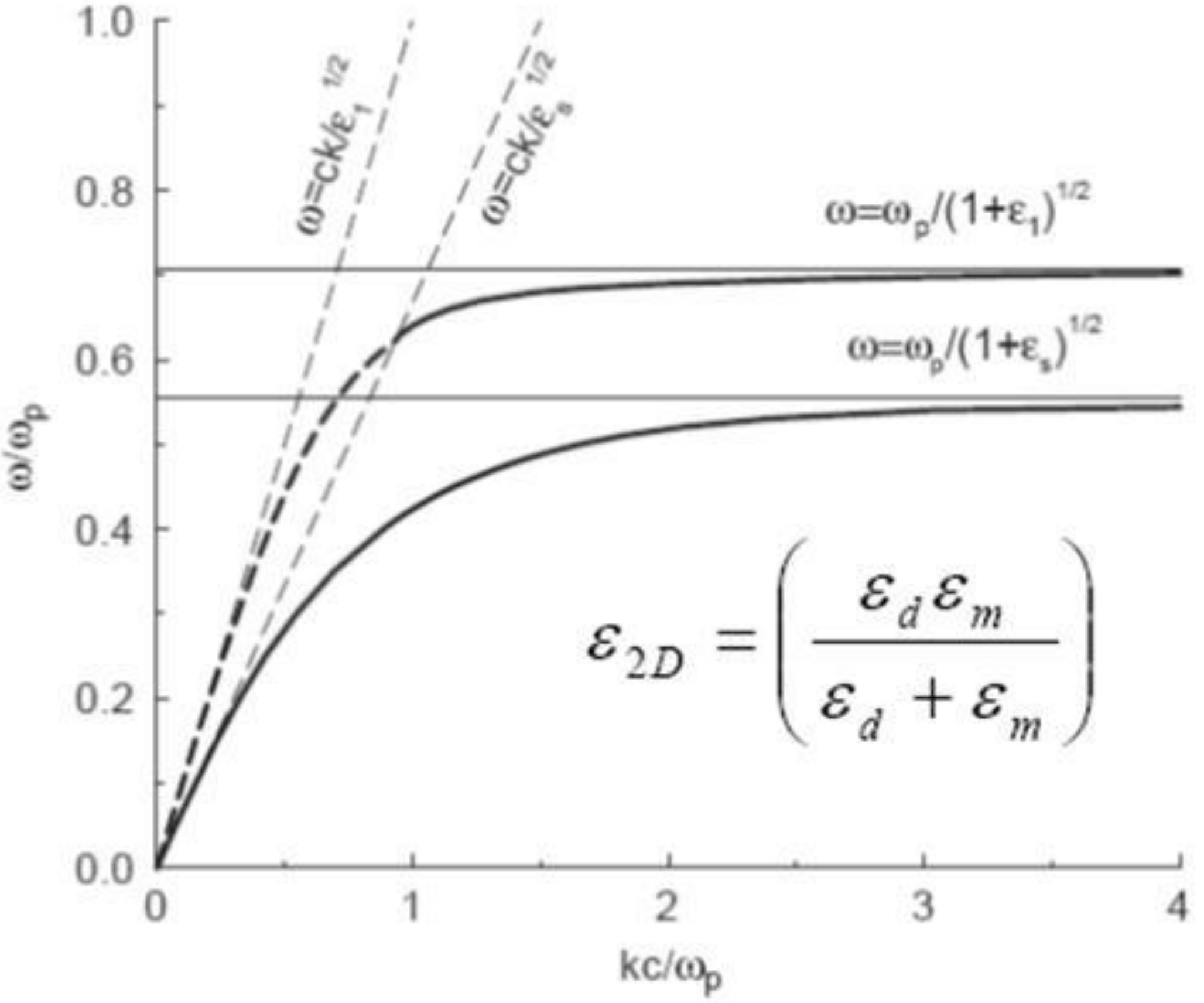

1.0
0.8
0.6
0.4
0.2
0.0
0
1
2
3
4
ω/ω_p
kc/ω_p
ω=ck/ε_1^1/2
ω=ck/ε_s^1/2
ω=ω_p/(1+ε_1)^1/2
ω=ω_p/(1+ε_s)^1/2
$\varepsilon_{2D} = \left( \frac{\varepsilon_d \varepsilon_m}{\varepsilon_d + \varepsilon_m} \right)$

Fig. 3

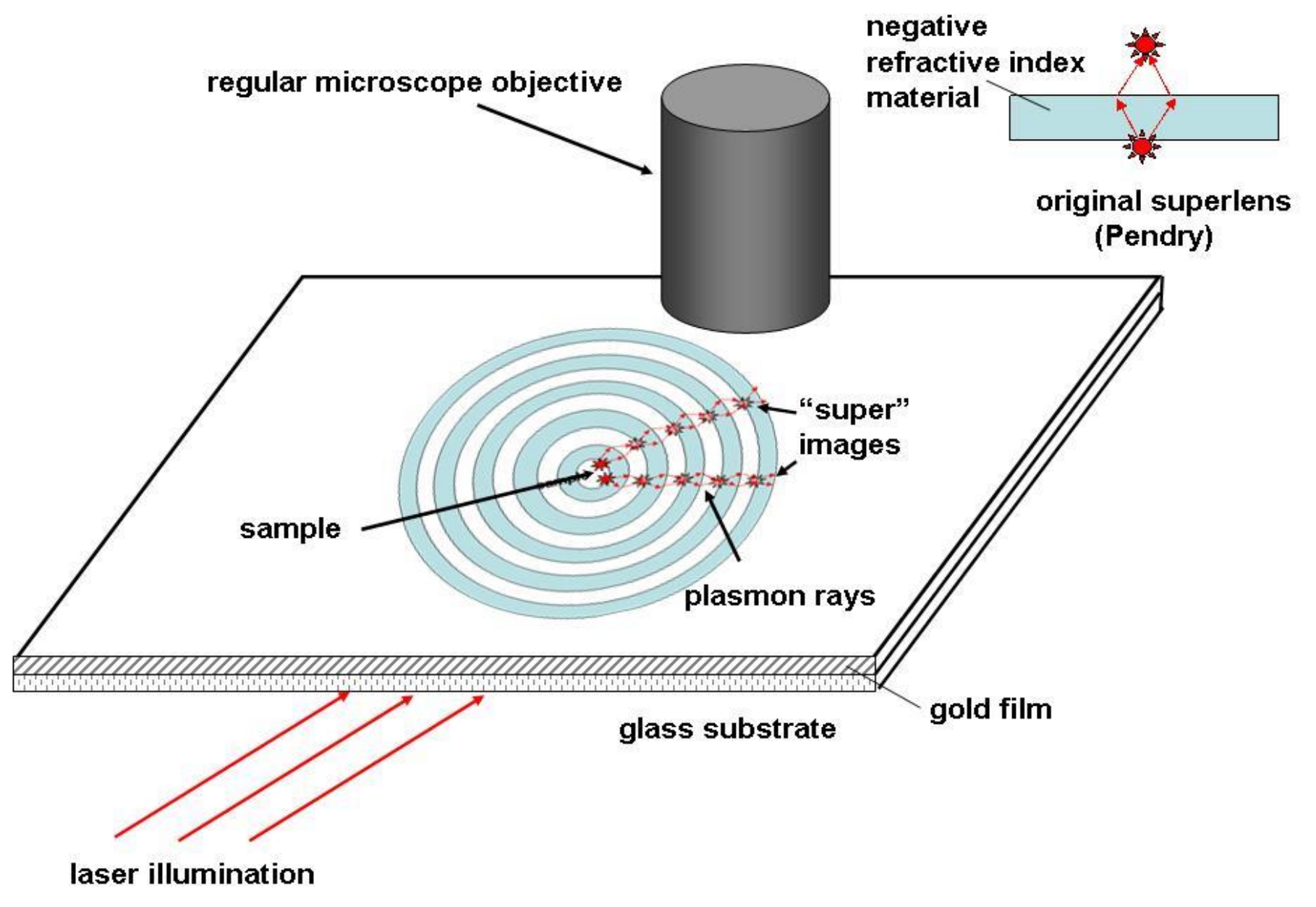

**(a)**

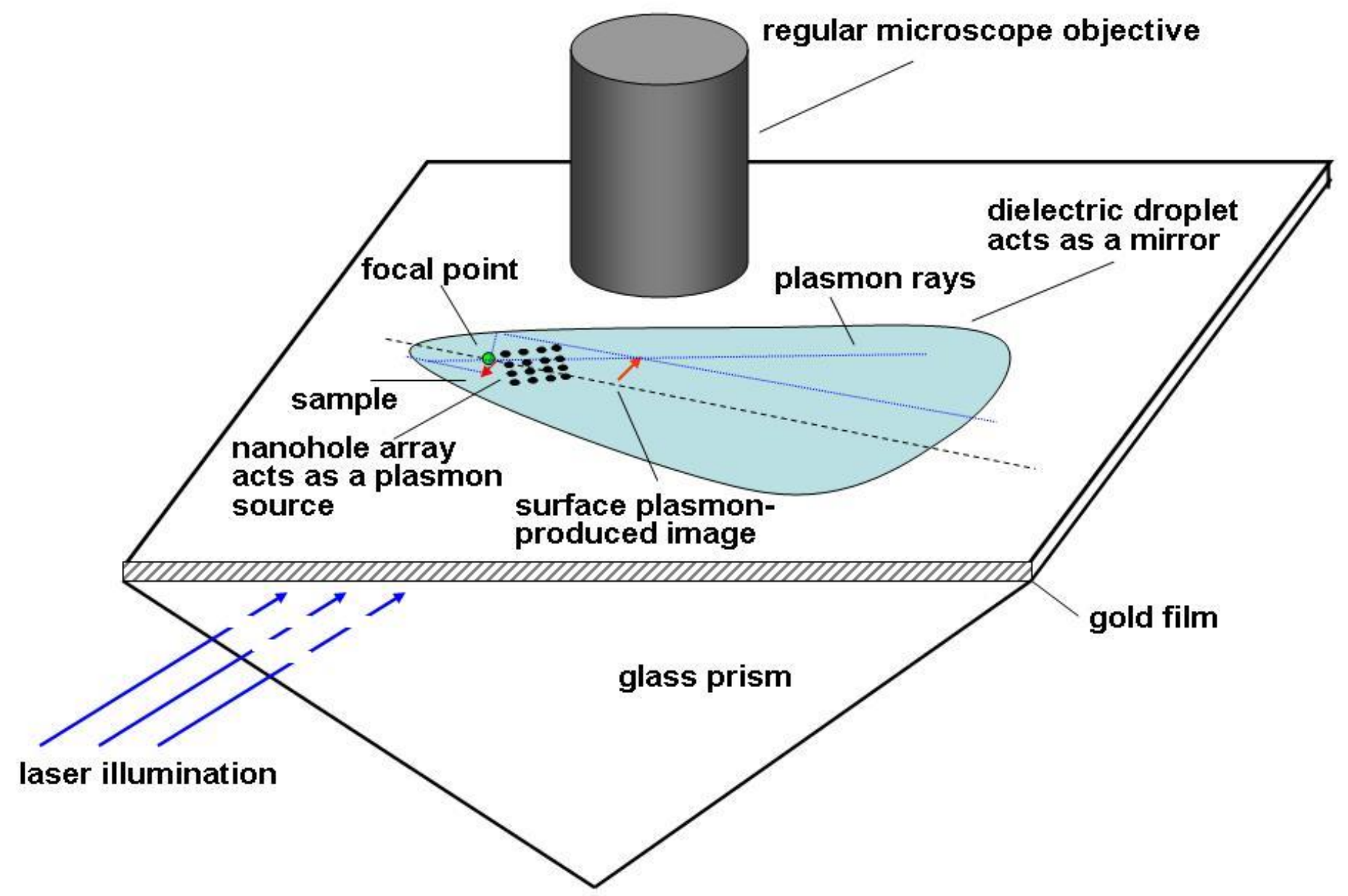

**(b)**

Fig. 4

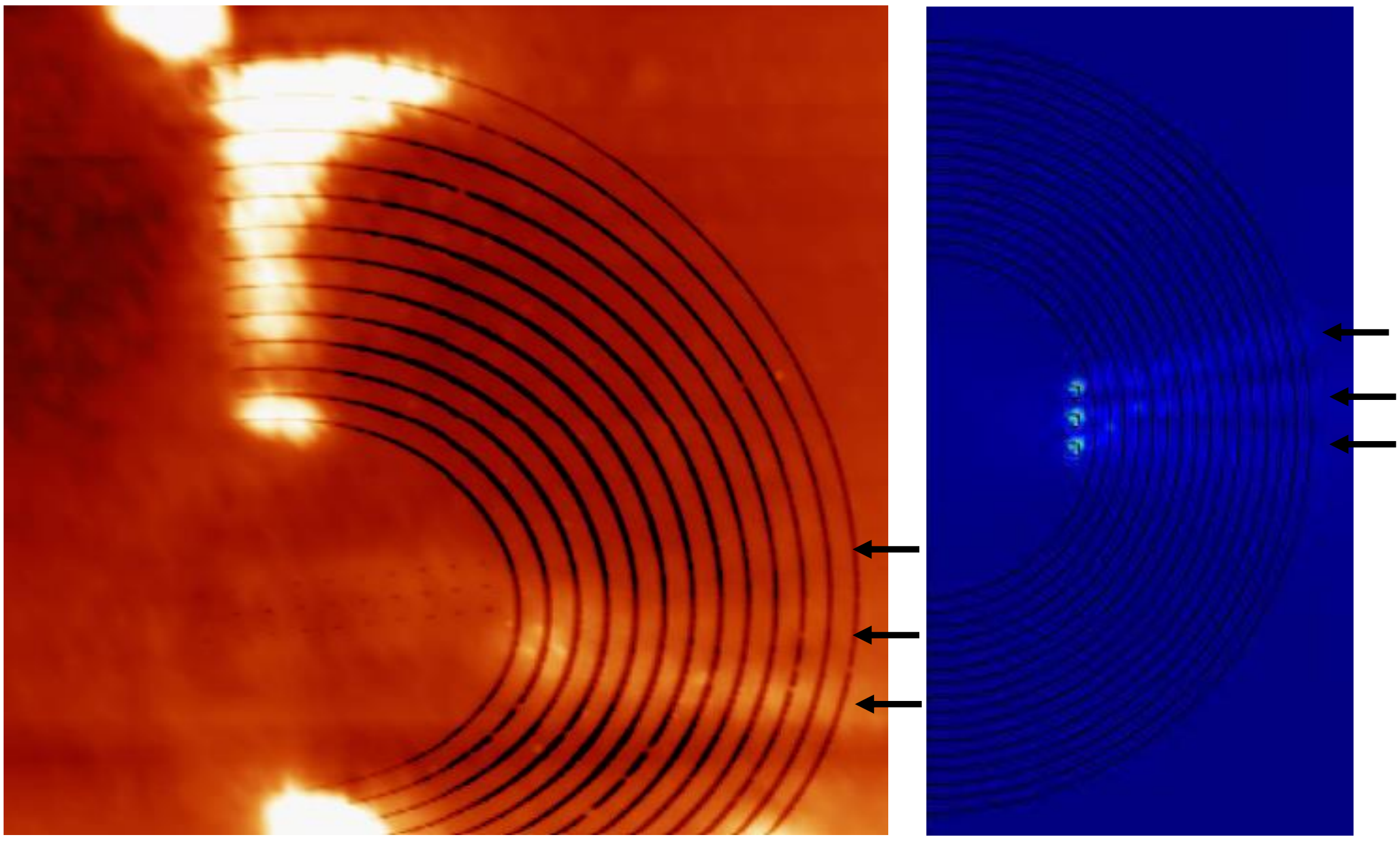

( a ) ( b )

Fig. 5

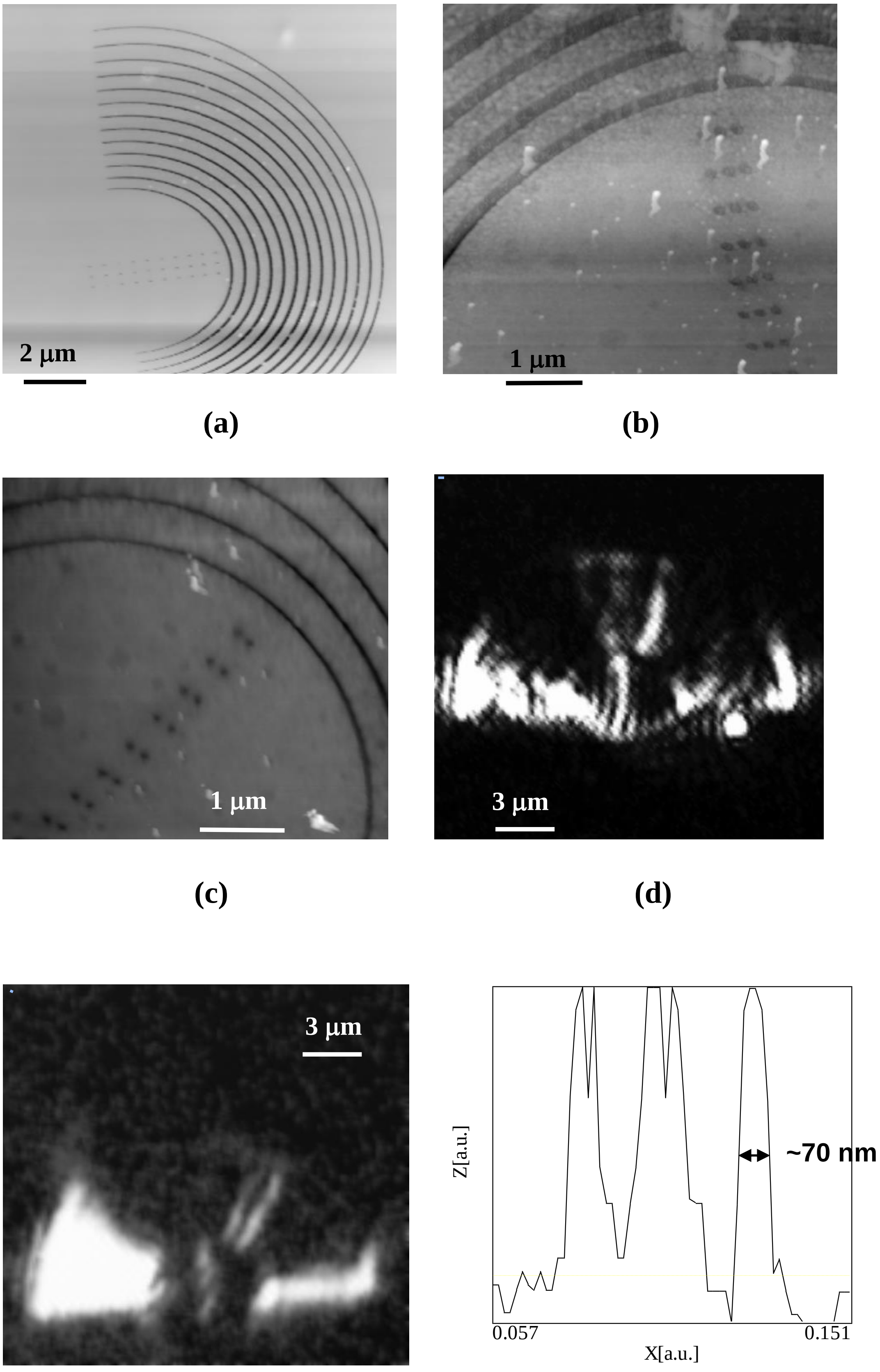

Fig. 6 (e) (f)

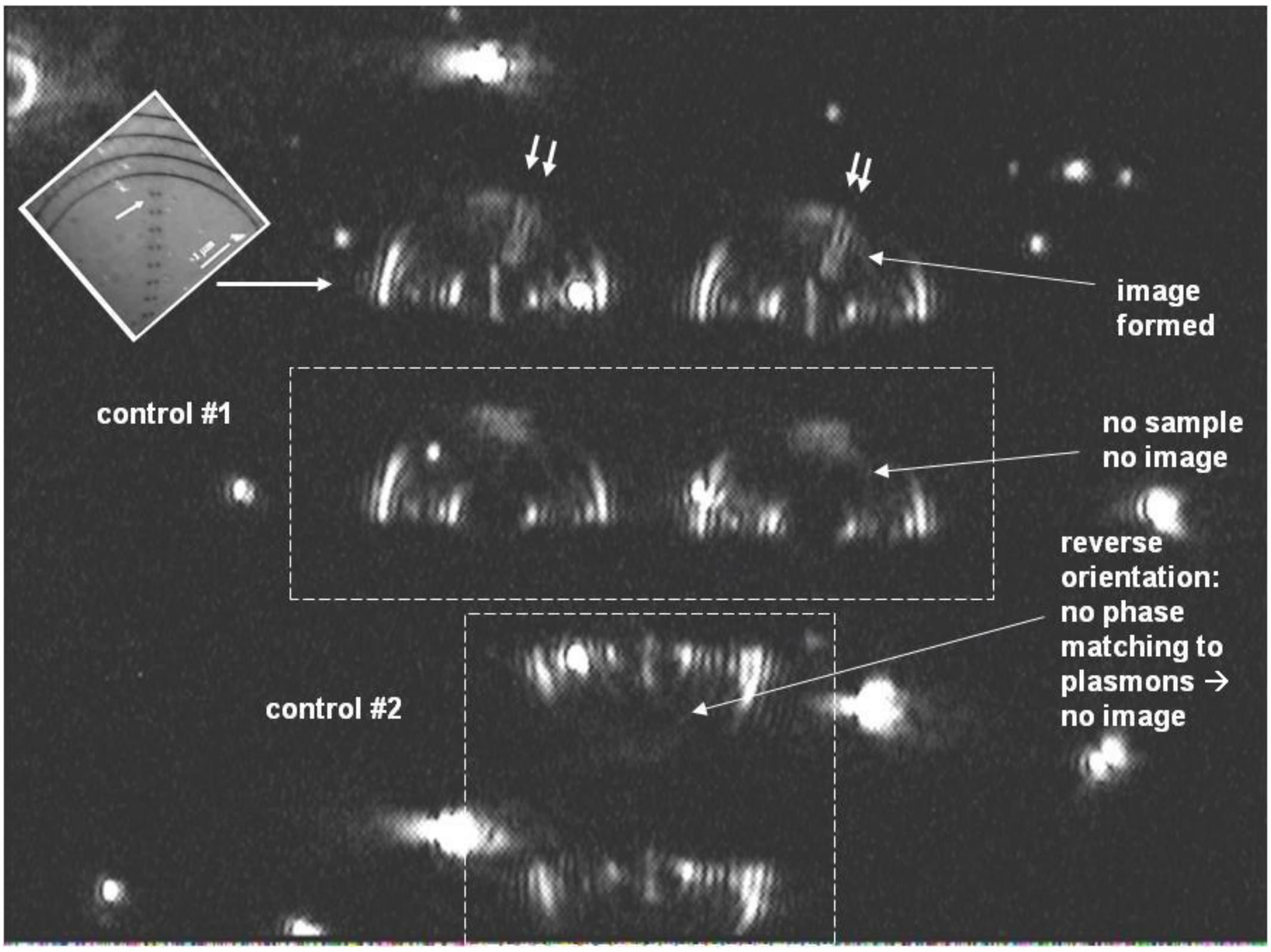

**(a)**

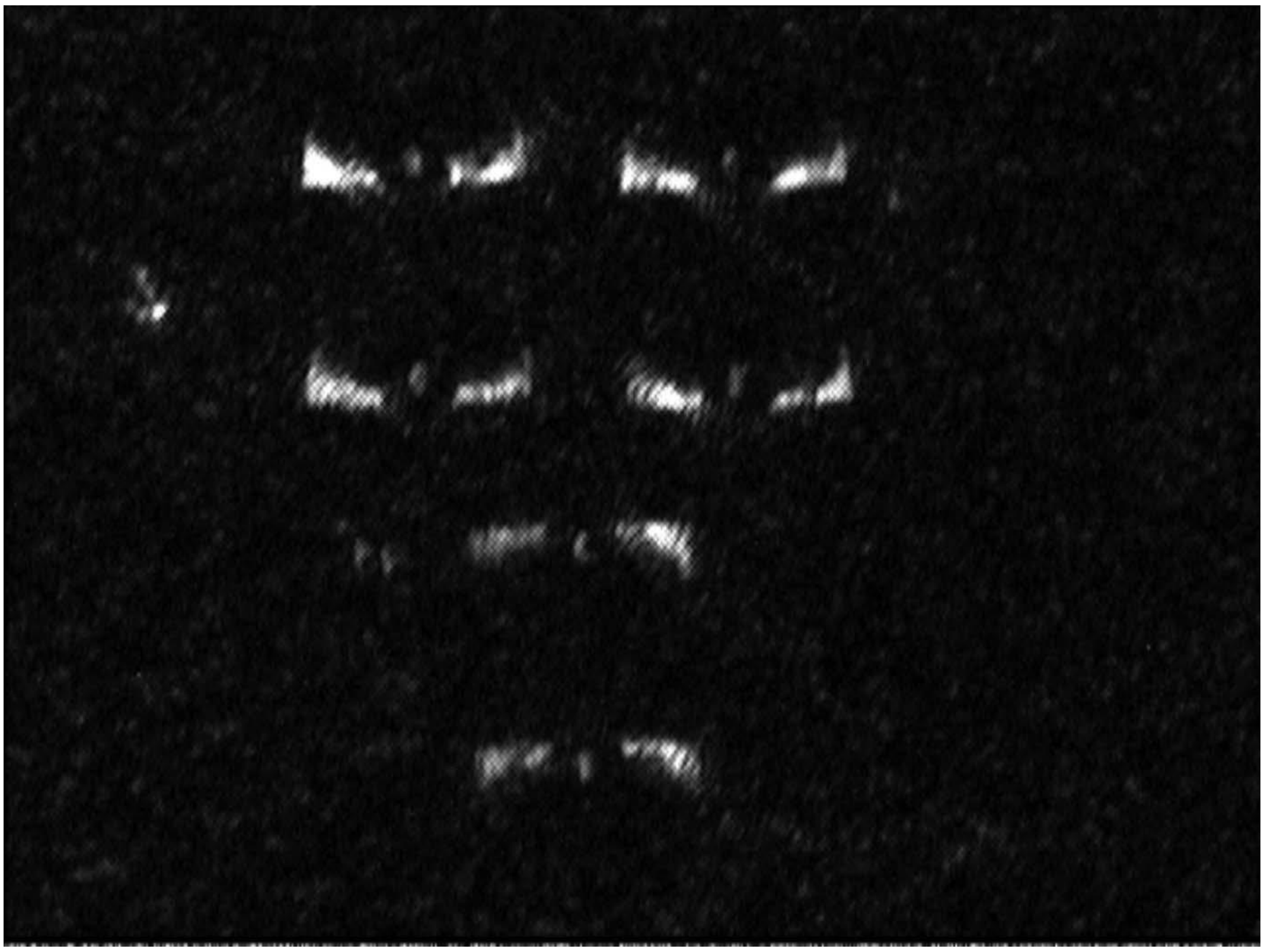

**(b)**

Fig. 7

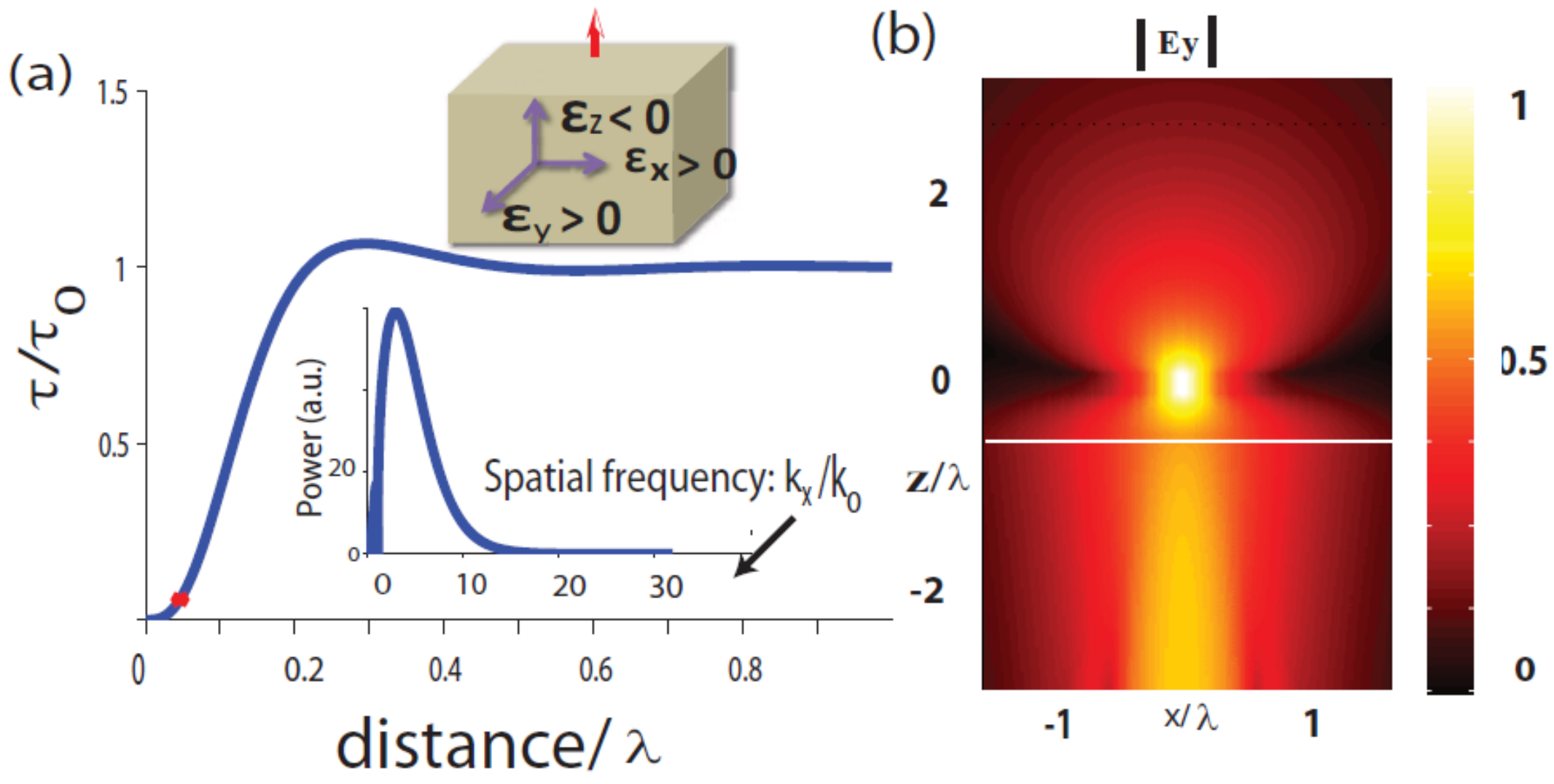

(a)
1.5
1
0.5
τ/τ0
εz < 0
εx > 0
εy > 0
Power (a.u.)
20
0
0
10
20
30
Spatial frequency: kx/k0
0
0.2
0.4
0.6
0.8
distance/ λ
(b)
|Ey|
2
0
-2
z/λ
-1
x/λ
1
1
0.5
0

Fig. 8

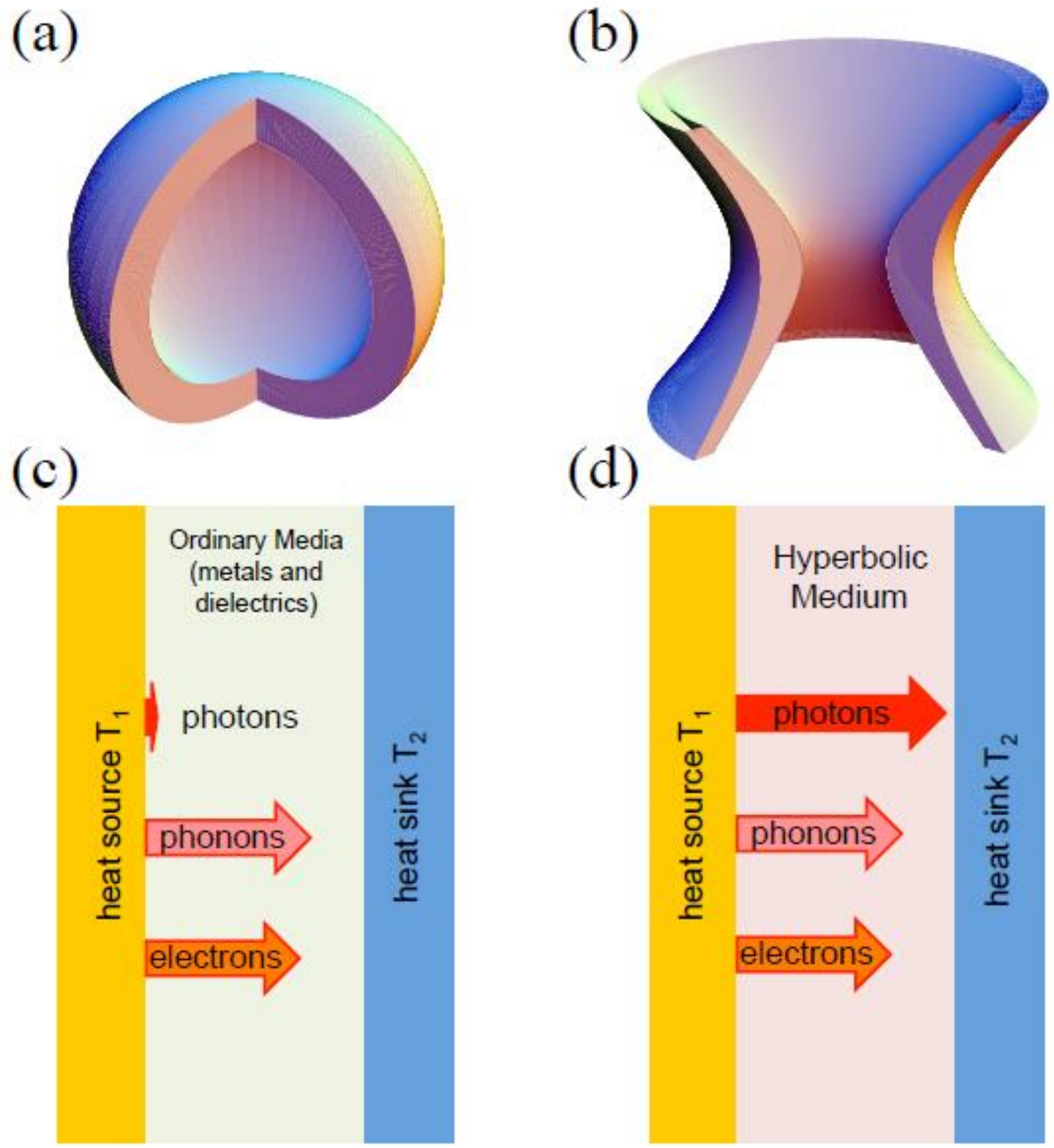

Fig. 9

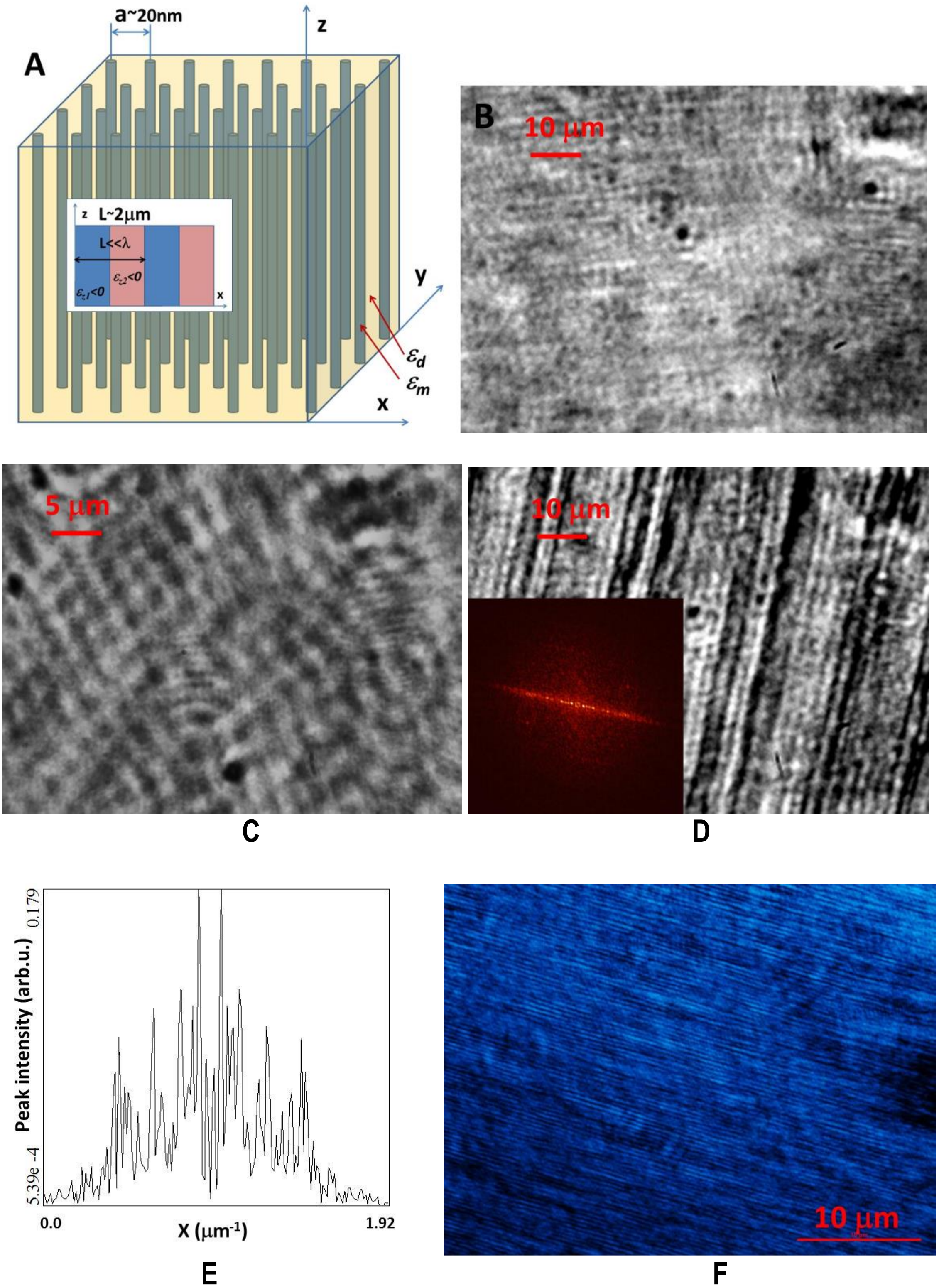

Fig. 10.

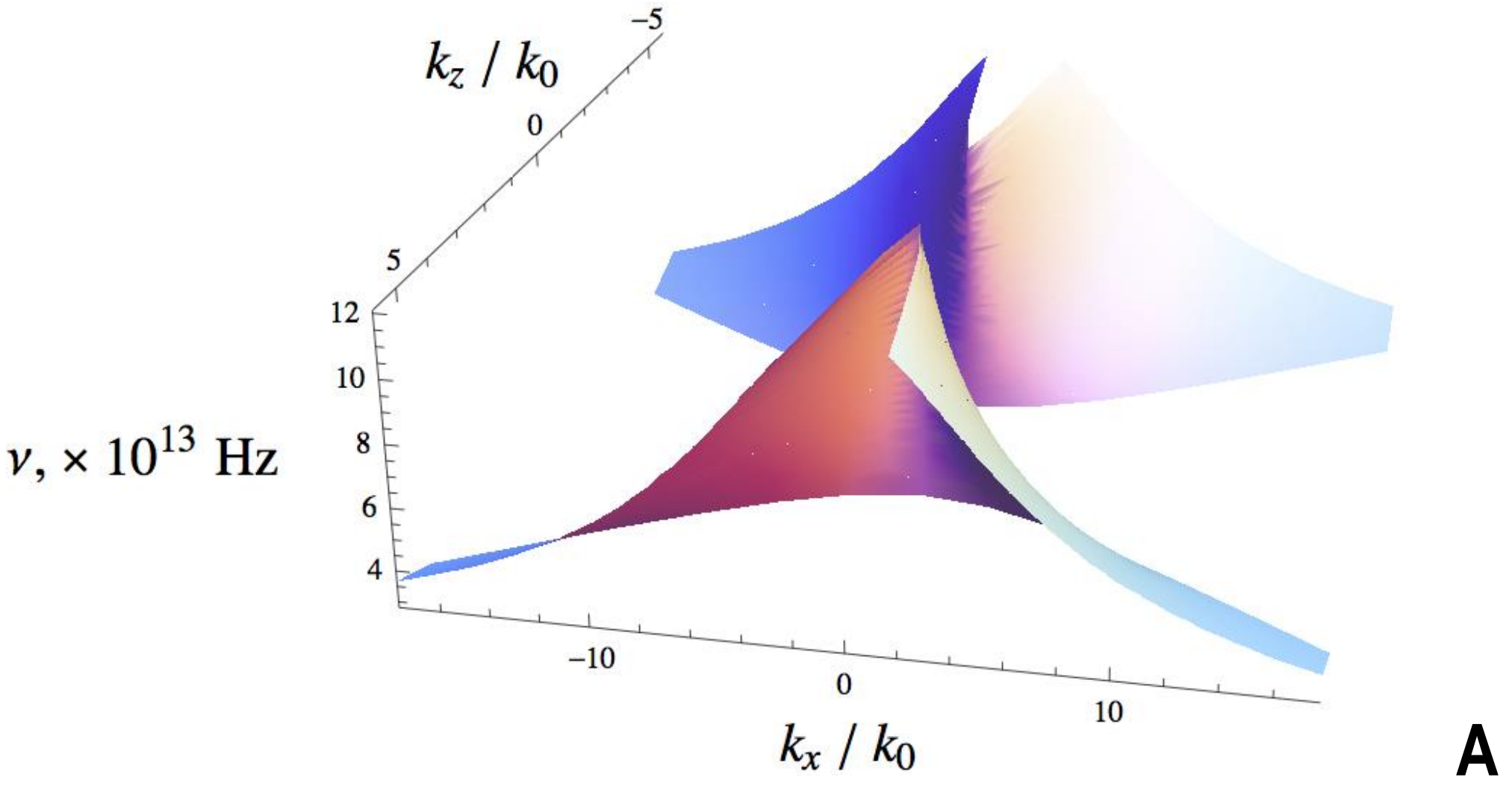

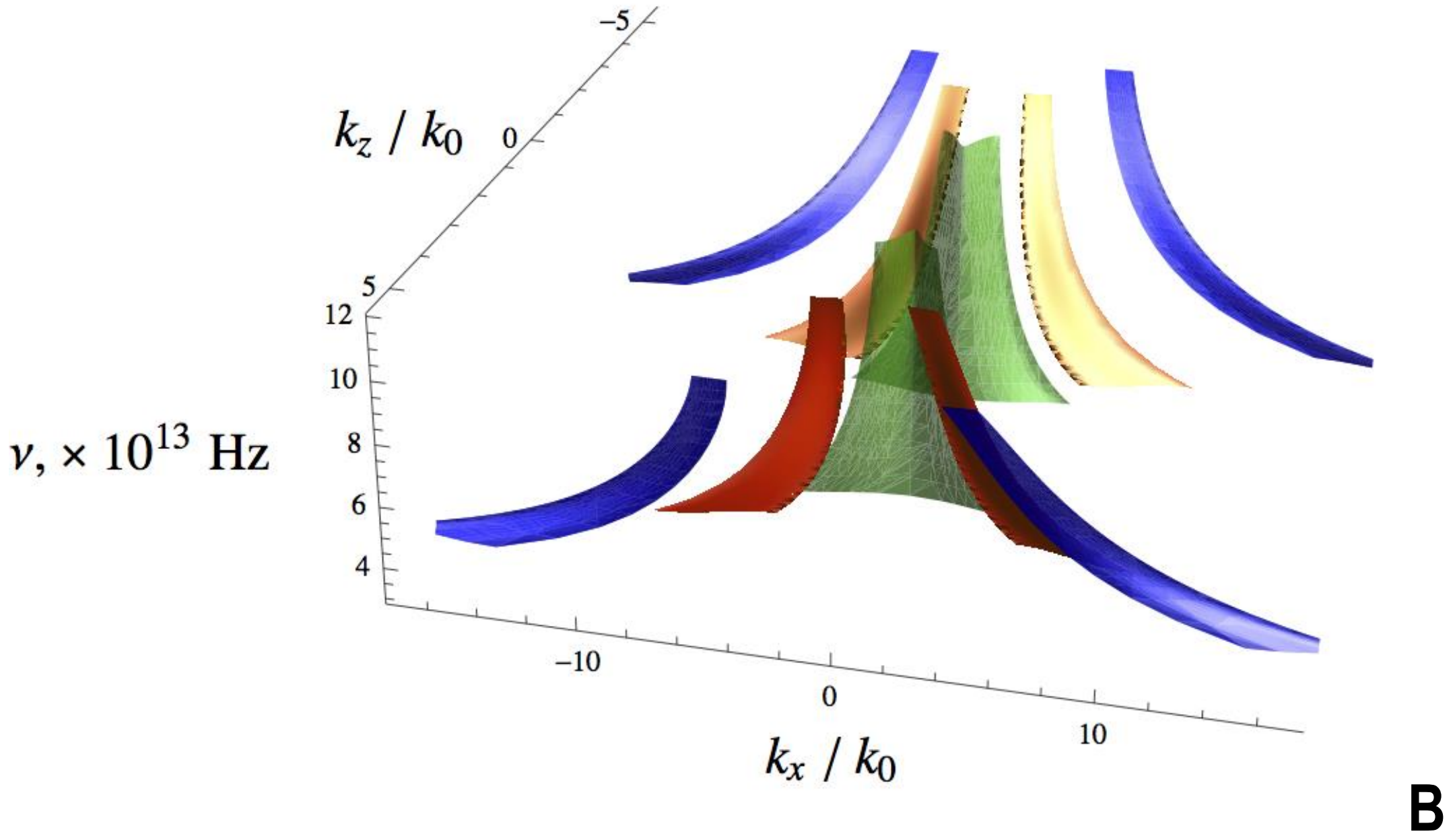

Fig. 11

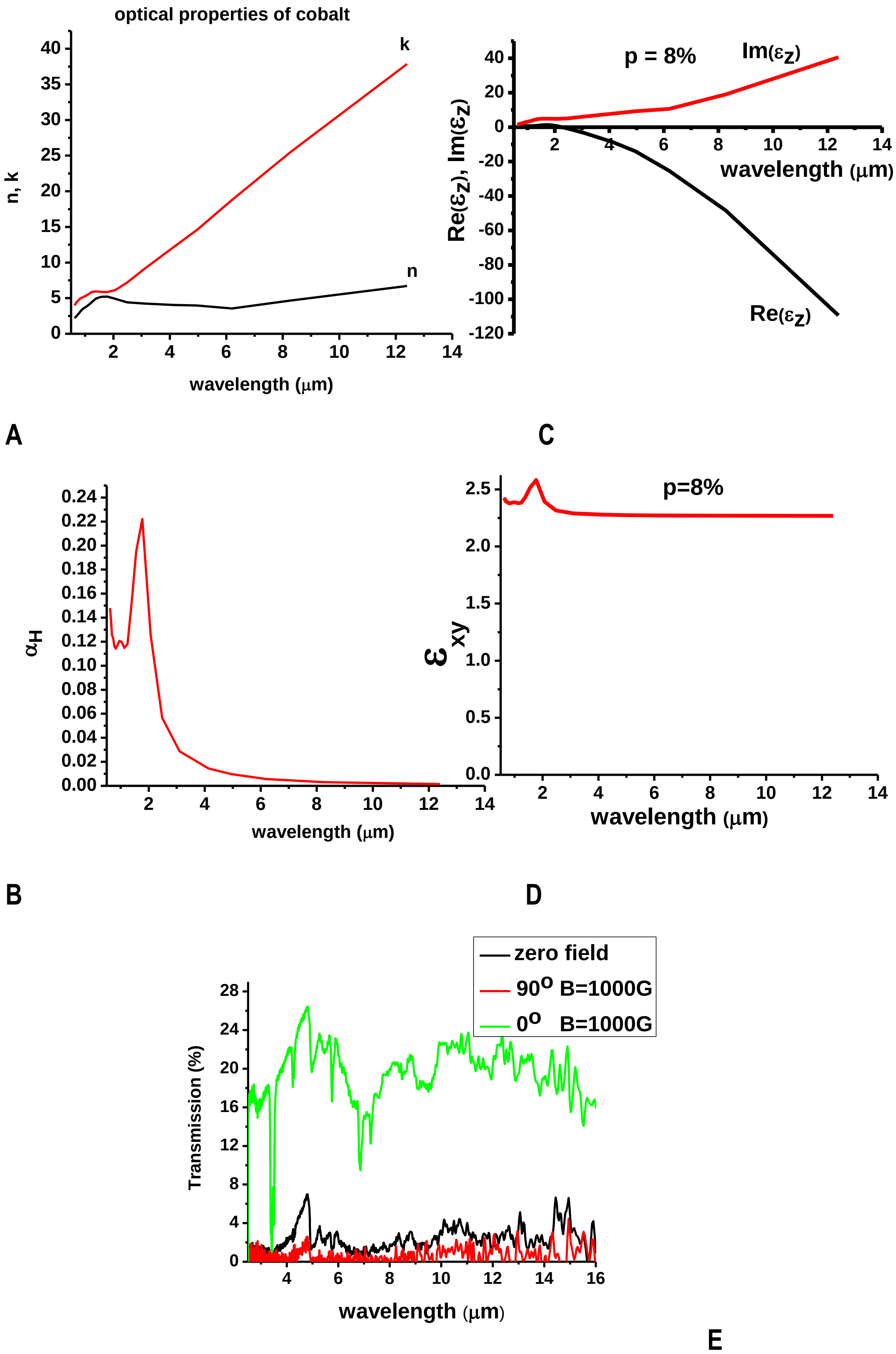

Fig. 12

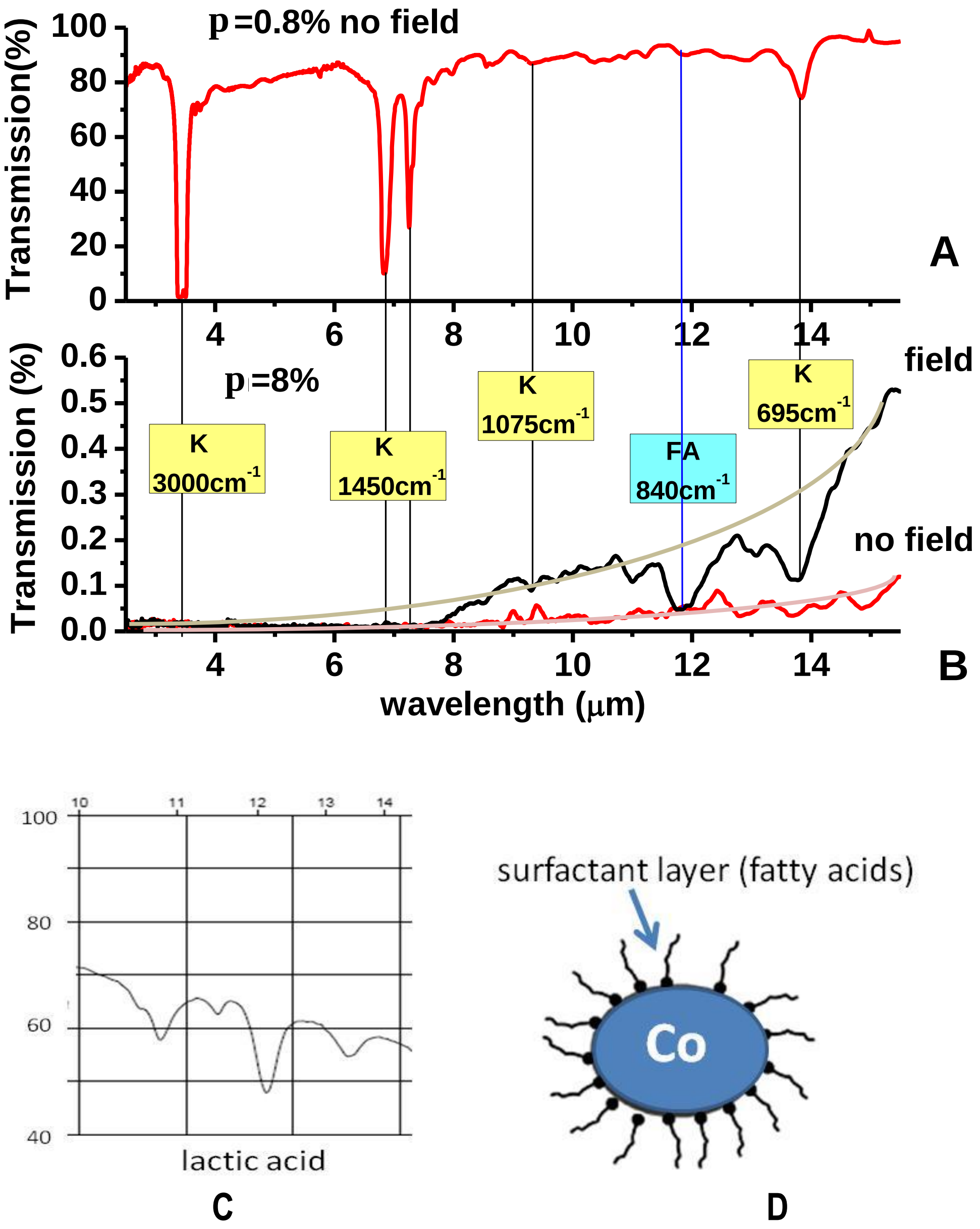

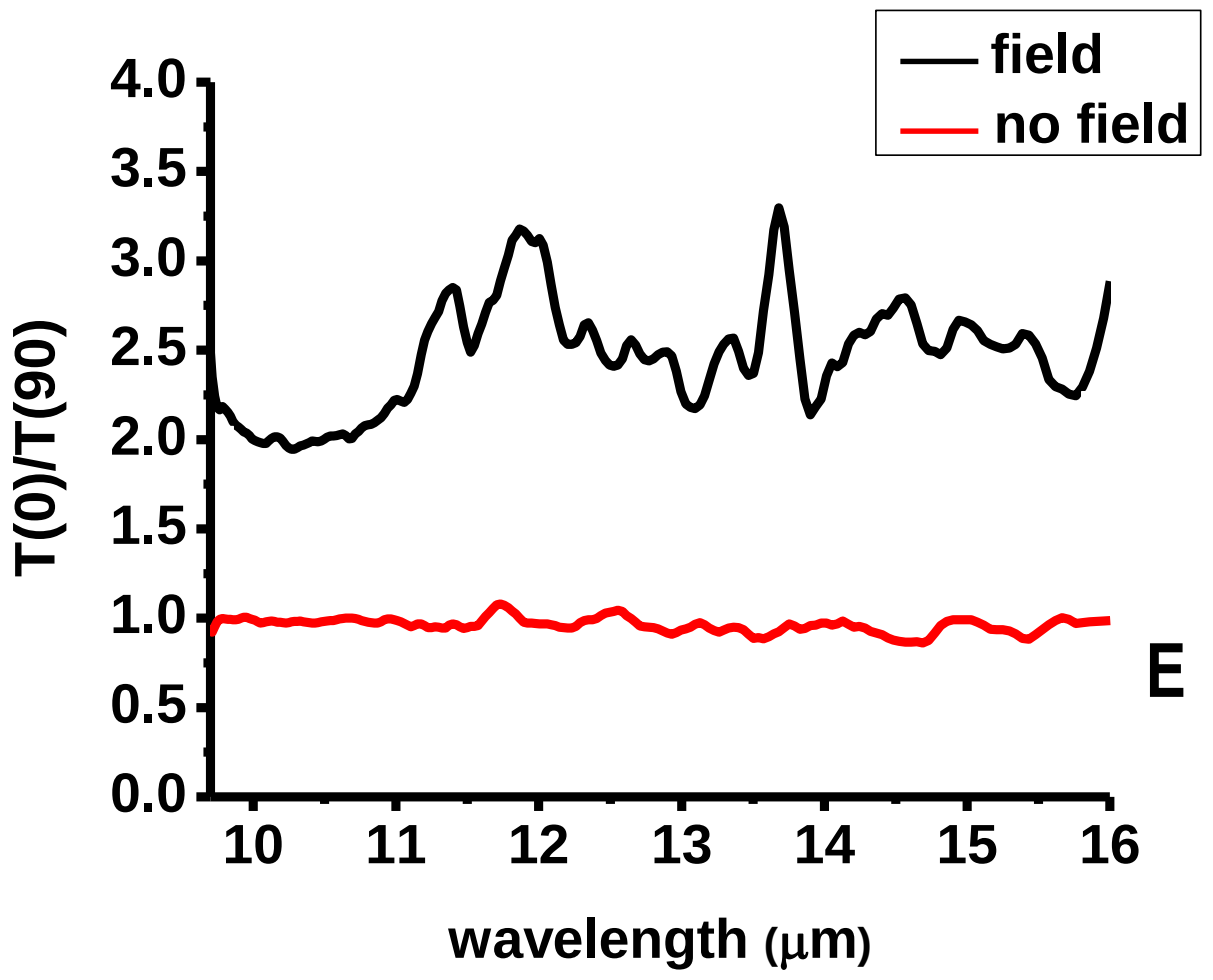

field
no field
T(0)/T(90)
E
wavelength (μm)

Fig. 13

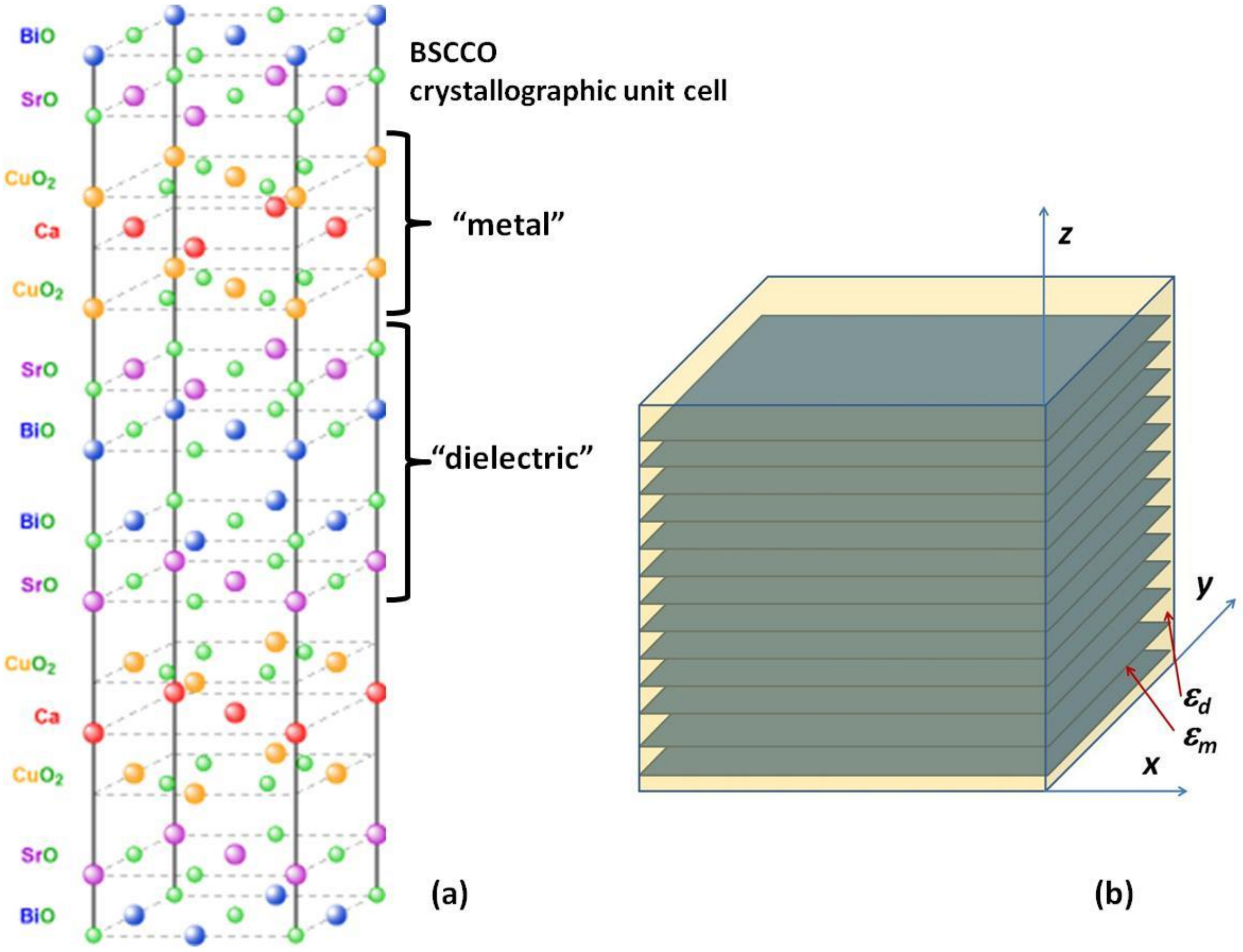

Fig.14

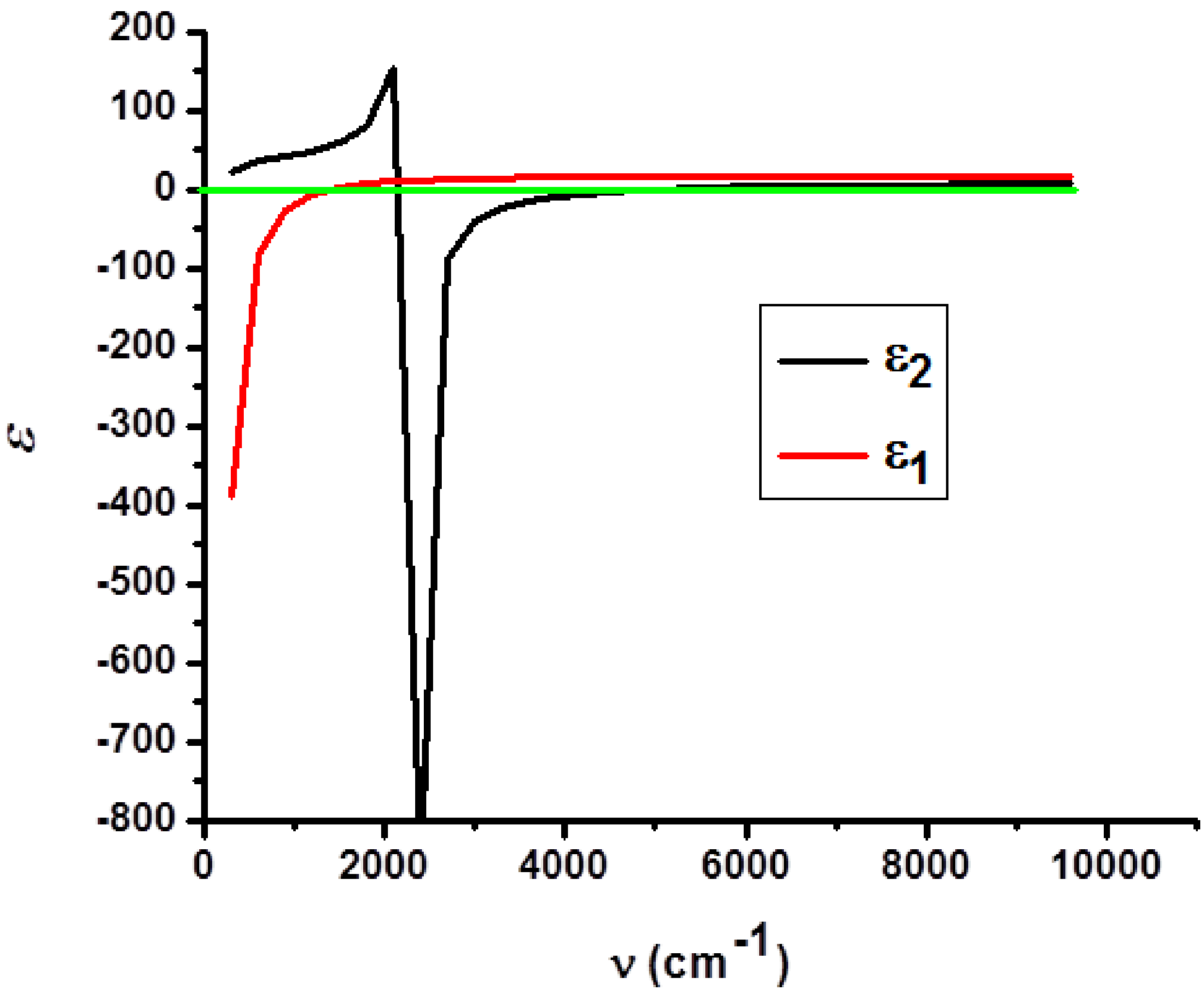

Fig. 15

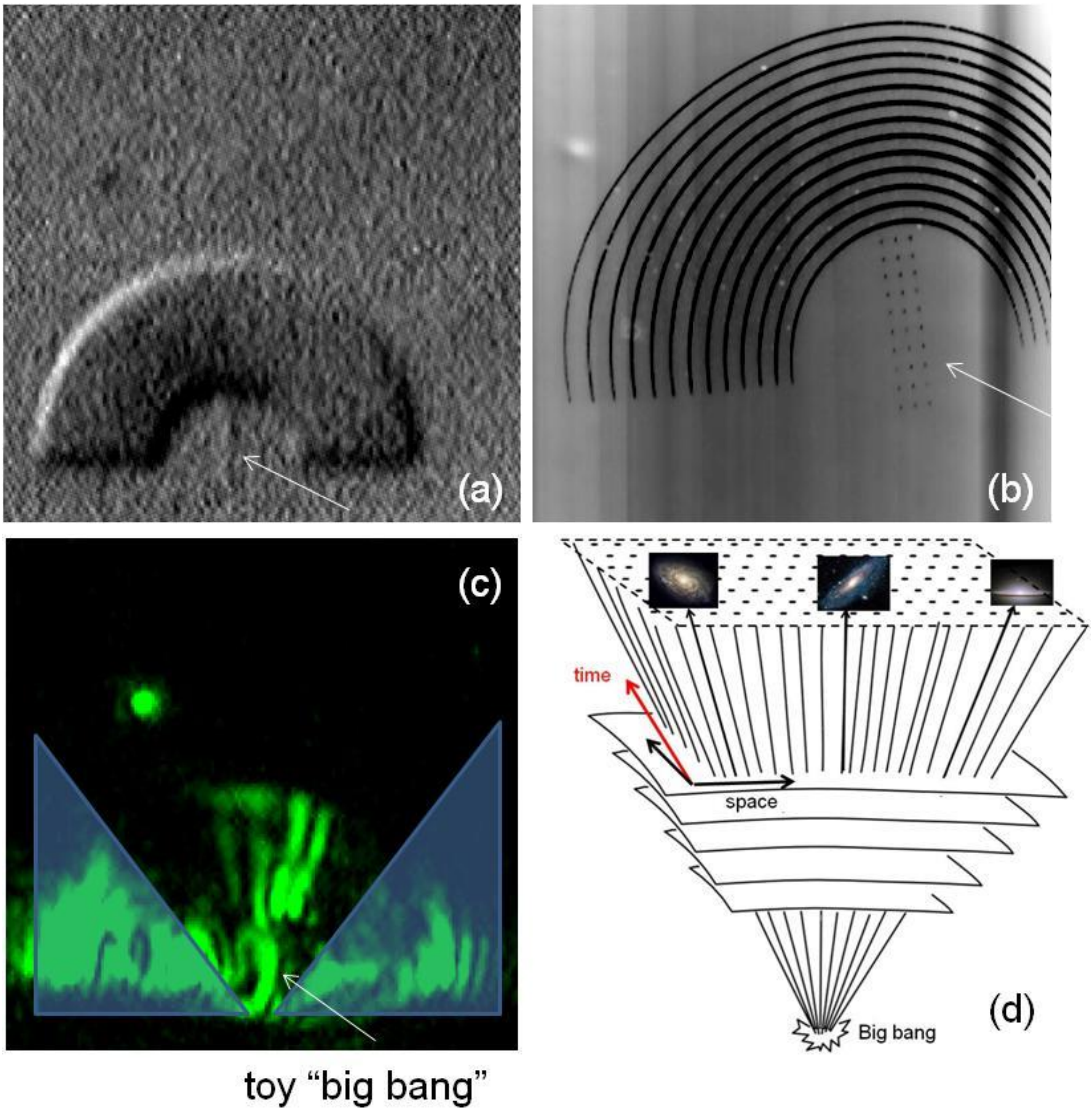

Fig. 16

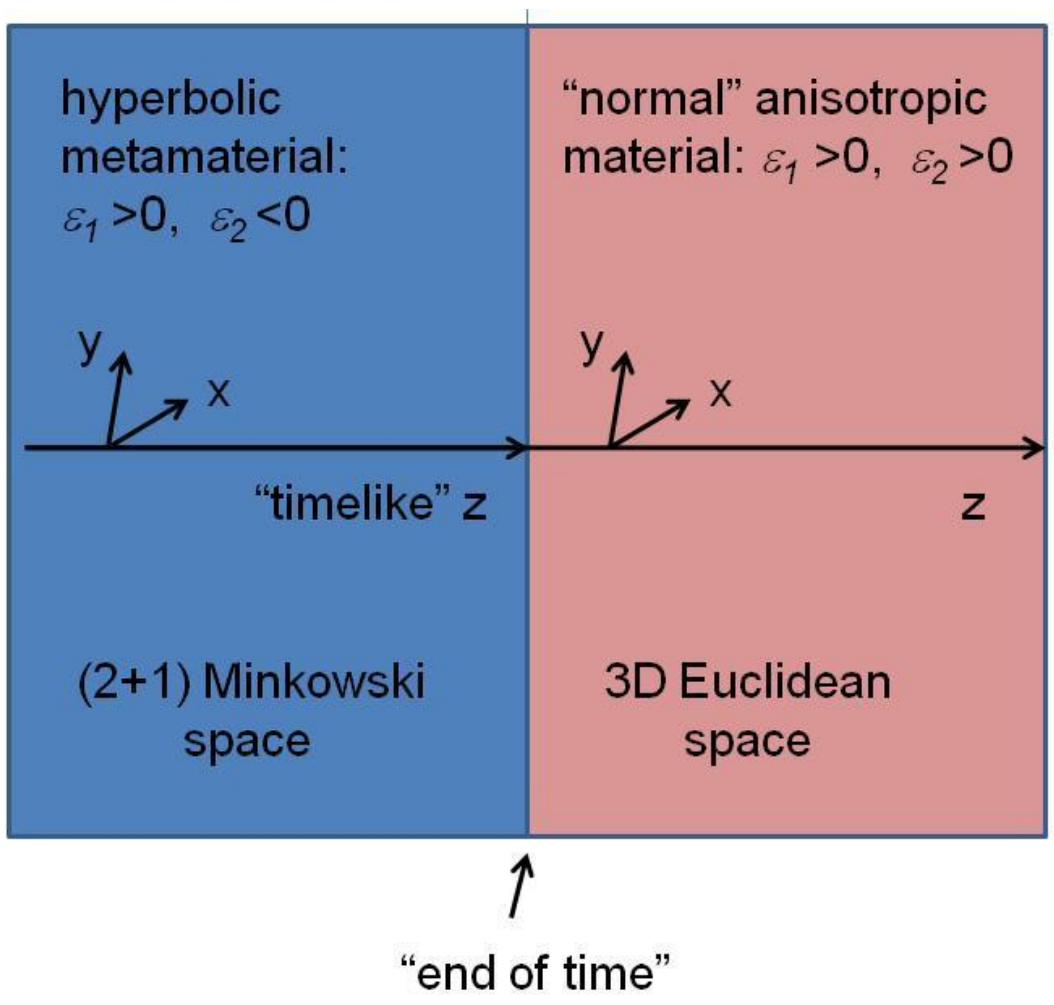
(a)

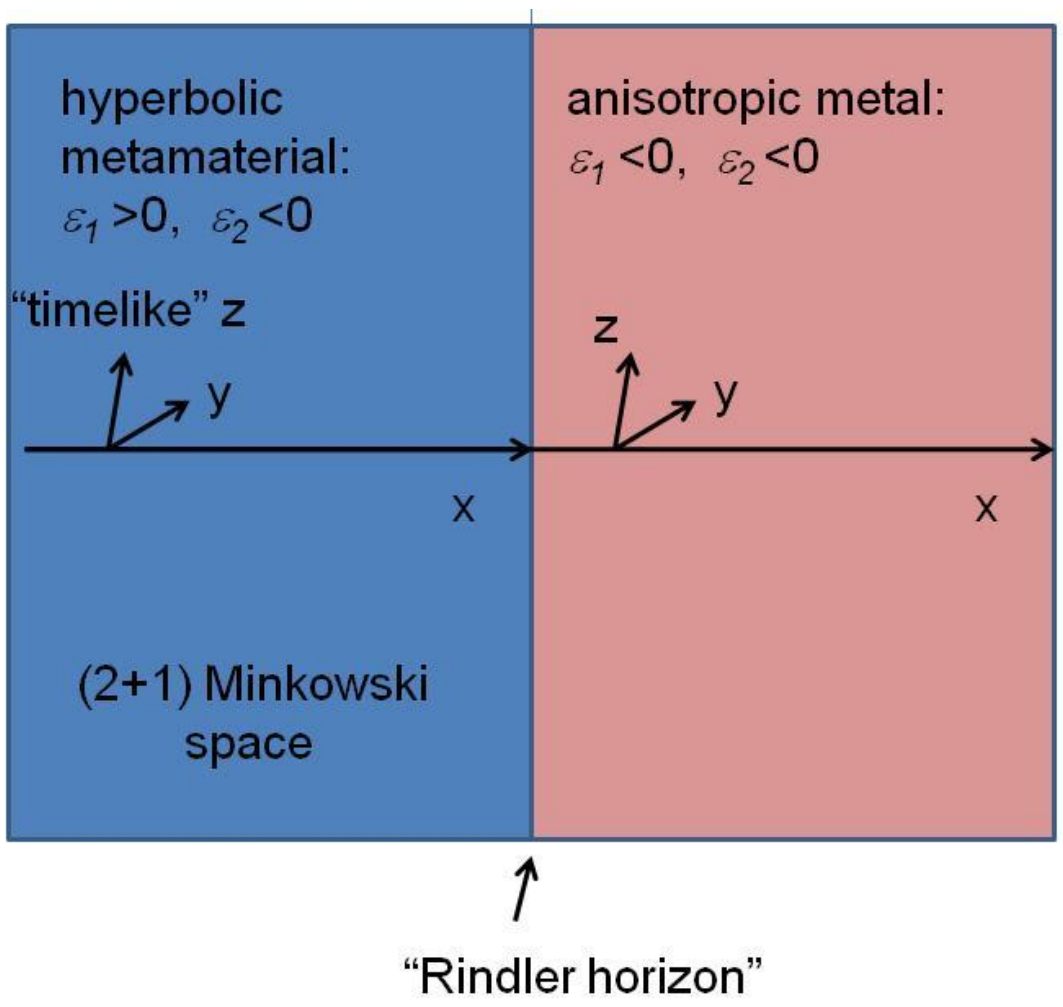
(b)

Fig. 17

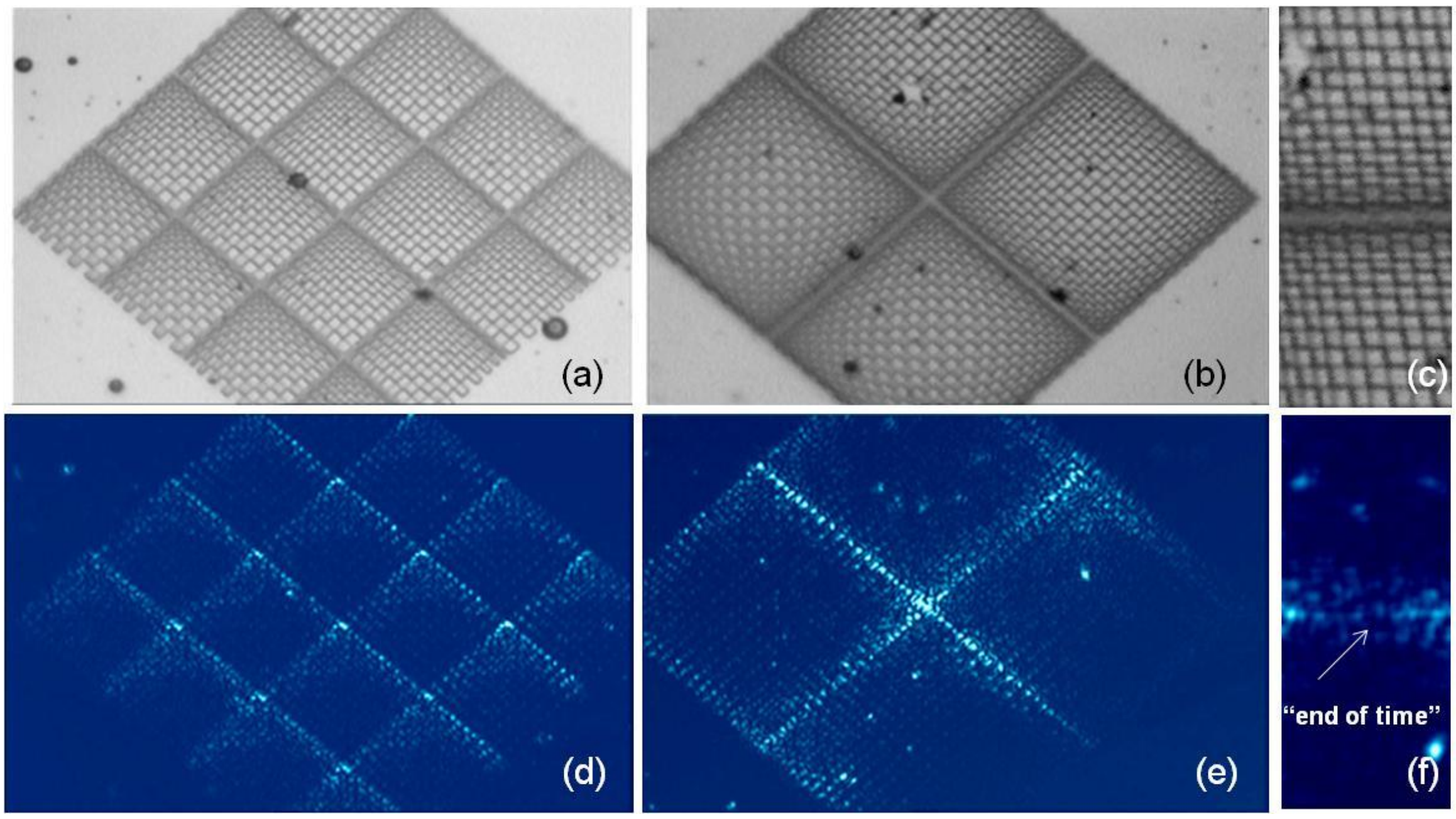

Fig.18

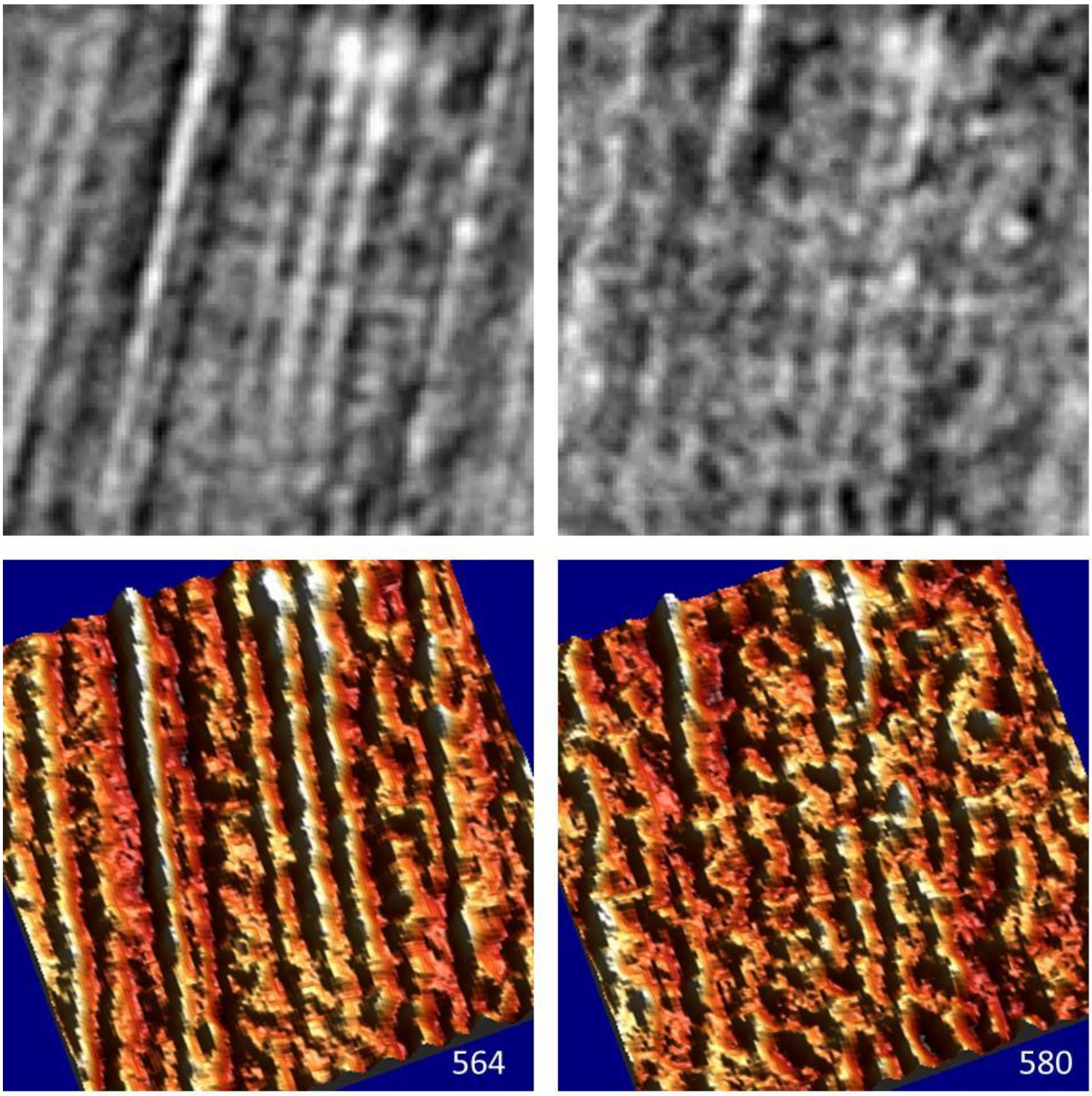

Fig. 19